\documentclass[UTF8,a4paper]{article}
\usepackage{ upgreek }
\usepackage{amsmath}
\usepackage{graphicx}
\usepackage{subfigure}
\usepackage{epstopdf}
\usepackage{geometry}
\usepackage{ulem}
\usepackage{indentfirst}
\usepackage{amsfonts,amssymb,dsfont}
\usepackage{setspace}
\usepackage{threeparttable}
\usepackage{amsmath,mathtools,amsthm}
\usepackage{array}
\usepackage{extarrows}
\usepackage{upgreek}

\makeatletter
\renewcommand*\env@matrix[1][\arraystretch]{%
  \edef\arraystretch{#1}%
  \hskip -\arraycolsep
  \let\@ifnextchar\new@ifnextchar
  \array{*\c@MaxMatrixCols c}}
\makeatother
\begin{document}


\title{\bf Bipolaron formed through electron-hole excitation}
\author{Chen-Huan Wu
\thanks{chenhuanwu1@gmail.com}
\\College of Physics and Electronic Engineering, Northwest Normal University, Lanzhou 730070, China}

\maketitle
\vspace{-30pt}
\begin{abstract}
\begin{large} 

We investigate the electronic properties and electron correlations 
of the bipolaron formed by the electron-hole excitations in the presence of Yukawa-type coupling
(between nonrelativistic fermions)
 in three spatial dimension.
The electron-hole excitation, which is necessary to the formation of bipolaron, 
leads to imaginary particle-hole order parameter,
and provide finite boson field mass to the single-polaron dispersion
in a broken-symmetry phase.
We found that the bipolaron exhibits fermi-liquid features as long as the long-range strong interaction is suppressed,
and it behave differently compared to the single-polaron.
The bosonic momentum determines the mass of boson field propagator and the gap function,
and it also related to the self-energies and the single-particle Green's functions.
The Thouless criterion is also used during the calculation of gap equation at critical temperature
(which become lower in weak-coupling regime),
which corresponds to the pole (instability) of the pair propagator in zero center-of-mass freamwork.
The mean field term and the bosonic fluctuation-induced contribution to free energy in Aslamazov-Larkin-type bipolaron diagram
are also studied.
\\

\end{large}

\end{abstract}
\begin{large}

\section{Introduction}

It has been examined that at weak polaronic coupling regime,
the mean-field treatment is valid in investigating the polaron energy as well
as its Fermi-liquid/non-Fermi-liquid behaviors.
While beyond the mean-field level, the order parameter fluctuations,
like the longitudinal propagator and the Goldstone propagator,
would leads to a relaxation rate which
is faster than that in the Fermi-liquid picture,
anomalously,
the critical Fermi surface may coexists with the long-lived stable quasiparticles,
and thus leads to unconvention transport properties.
Such case usually related to the scalar bosonic degrees of freedom
and the quantum critical points like the charge density wave states,
whose order parameter is a scalar.
Examples includes the transition from Landau damping to diffusive one by the disorder
and accompanied by a quantum phase transition.

For inclosed diagrams of bipolaron as a collective mode (in contrast to the closed diagrams of the Aslamazov-Larkin-type bipolaron),
the self-energy $\Sigma$ and spectral function $A$ has been calculated and discussed,
which exhibit that the bipolaron is in fermi-liquid state at least at low-temperature,
where $|{\rm Im}\Sigma(\mathcal{W})|\ll |\mathcal{W}|$ ($\mathcal{W}\rightarrow 0$),
and the resonance pole (corresponds to the the bound state) appears at negative $\mathcal{W}$.
The non-fermi-liquid behavior is no shown since, 
firstly, the long-range interaction is suppressed (screened to the instantaneous) by the 
particle-hole excitations (due to the quasiparticle spectral weight at extended fermi surface especially at high spatial dimensions),
and secondly, the quantum fluctuations is suppressed even in the low energy limit by the
massive bosonic modes.
It is worth to note that, the non-fermi-liquid phase with weak singularity at $Z=0$ fermi surface\cite{Abrahams E}
would be found in strong coupling regime where the quantum critical behavior can be 
extends to high energy scales, and the relation $|{\rm Im}\Sigma(\mathcal{W})|< |\mathcal{W}|$ is still satisfied
and the spectral function can still be sharply peaked.

\section{Model}

Firstly we write the Hamiltonian of our bipolaron model as
\begin{equation} 
\begin{aligned}
H=&\sum_{p_{1}}\varepsilon_{p_{1}}\psi^{\dag}_{p_{1}}\psi_{p_{1}}+\sum_{p_{2}}\varepsilon_{p_{2}}\psi^{\dag}_{p_{2}}\psi_{p_{2}}
+\sum_{k}\varepsilon_{k}\psi^{\dag}_{k}\psi_{k}\\
&+g_{\psi\phi}\sum_{p_{1},k,q}\psi^{\dag}_{p_{1}-q}\psi^{\dag}_{k+q}\psi_{k}\psi_{p}
+g_{\psi\phi}\sum_{p_{2},k,q}\psi^{\dag}_{p_{2}+q}\psi^{\dag}_{k}\psi_{k+q}\psi_{p_{2}}.
\end{aligned}
\end{equation}
There are the four point vertices in the second line of above formula,
which is widely seen in the $T$-matrix approximation\cite{1,2,3,4,5}.
Note that in many-particle generalization,
we have the average $\langle \psi^{\dag}_{p_{1}-q}\psi^{\dag}_{k+q}\psi_{k}\psi_{p}\rangle
=\langle \psi^{\dag}_{p_{1}-q}\psi_{p}\rangle \langle \psi^{\dag}_{k+q}\psi_{k}\rangle
-\langle \psi^{\dag}_{p_{1}-q}\psi_{k}\rangle \langle \psi^{\dag}_{k+q}\psi_{p}\rangle
+\Pi(p_{1},k)$, where $\Pi(p_{1},k)$ is the pair propagator.
In this expression,
the first two terms give the Hartree-Fock mean-field energies, while the last term gives the correlation energy.
This expression can also be writen in terms of the two-particle Green's function
(second functional derivative of the partition function), see Ref.\cite{Haussmann R}.
This four fermion field vertex can be turned to the Yukawa coupling by using the Hubbard-Stratonovich transformation.
Then base on the above Green's functions (in frequency domain),
we can write the action of bipolaron system $S_{g}$ (related to Yukawa coupling) as
\begin{equation} 
\begin{aligned}
S_{g}=g_{\psi\phi}\int\frac{d^{3}p}{(2\pi)^{3}}\frac{d^{3}q}{(2\pi)^{3}}
\overline{\psi}(p-q)\psi(p)\phi(q)
+g_{\psi\phi}\int\frac{d^{3}k}{(2\pi)^{3}}\frac{d^{3}q}{(2\pi)^{3}}
\overline{\psi}(p)\psi(p-q)\phi^{*}(q),
\end{aligned}
\end{equation}
where $\psi$ is the Fermionic Grassmann field and the flavor factor
$1/\sqrt{N}$ is omitted in each term (each single interaction vertex).
Here
$\phi(q)\neq \phi^{*}(q)$ is guaranteed by the particle-hole asymmetry.
This action describes the lowest-order coupling between bosonic order-parameter fluctuations
and the fermions.
$\overline{\psi}(p_{i})=\psi^{\dag}\gamma^{0}$ is necessary here
as long as $\psi$ is not a complex scalar field,
to satisfy the Lorentz invariant condition.
$\overline{\psi}$ and $\psi$ can be treated as
algebraically independent Grassmann variables
to make it accessible to vary the action
with respect to both of them separately.
$\phi(q)$ is the scalar order parameter field (bosonic degrees of freedom) in particle-hole channel,
which reads
\begin{equation} 
\begin{aligned}
\phi(q)=ig_{\psi\phi}\int\frac{d^{3}k}{(2\pi)^{3}}
  \overline{\psi}(k+q)\psi(k).
\end{aligned}
\end{equation}
$g_{\psi\phi}(<0)$ 
denotes the coupling strength between fermion field and scalar bosonic field (in the vertices
of Yukawa coupling), which also plays the role of symmetry factor.
The existence of $g_{\psi\phi}$ here reveals that the coupling between bare boson field $\phi(q)$ and the particle-hole bubble
is equivalent to the Yukawa coupling,
and it is thus direct that the order parameter field in Eq.() should not contains the symmetry factor again as
the four-point vertices in Eq.(1) does not have a squared factor $g^{2}_{\psi\phi}$.
That is to say, the order parameter simply has a bilinear fermionic form\cite{Veillette M Y} in Eq.(1).
Note that in particle-particle channel the order parameter has a different form with the above one,
see Refs.\cite{Dong X,Macridin A,Hague J P,Strack P}.
In the presence of four fermions interaction vertex in Hubbard-Stratonovich space,
we suppose these interactions stabilize the charge density fluctuations.
$\phi(q)$ here provides a mass term $m_{\phi}$ to the fermionic quasiparticle dispersion,
i.e., the polaron dispersion,
as
$\varepsilon'_{p}=\sqrt{\varepsilon_{p}^{2}+|m_{\phi}|^{2}}+\Sigma(p,\omega)$,
where $\Sigma(p,\omega)$ is the normal self-energy 
(i.e., the diagonal element of the polaron self-energy matrix) induced by the polaronic coupling between impurity
and majority particles.
Combined with the above-mentioned non-diagonal static bosonic self-energy,
we can have the fermion self-energy matrix as
$\Sigma=\begin{pmatrix}
\Sigma(p,\omega) & \Delta\\
\Delta^{*}  & -\Sigma(-p,-\omega)
\end{pmatrix}$.
The gap term can be $\Delta$ can be divided into
the fluctuation part and mean-field part (the anomalous self-energy),
at low-temperature with finite order-parameter fluctuation.
The mean-field approximation here corresponds to the limit of large spatial dimensions\cite{Georges A},
and ignores the spatial fluctuations induced by nonlocal impurity.
The opening of mass gap in quasiparticle spectrum by the condensation of the bosonic field
also observed in the Dirac or semi-Dirac systems\cite{Christou E,Janssen L,Uryszek M D}
and the BCS gas (below BCS critical temperature)\cite{Ohashi Y}.
Thus the resulting fermion field Green's function of the single-particle excitation
reads
\begin{equation} 
\begin{aligned}
G_{p,\omega}
=&\frac{1}{i\omega-(\varepsilon_{p}+\Delta)}\\
=&-\frac{i\omega+(\varepsilon_{p}+\Delta)}{\omega^{2}+(\varepsilon_{p}+\Delta)^{2}},
\end{aligned}
\end{equation}
which in imaginary time domain reads
\begin{equation} 
\begin{aligned}
G_{p,\tau}
=&\int\frac{d\omega}{2\pi}\frac{e^{-i\omega\tau}}{i\omega-(\varepsilon_{p}+\Delta)}\\
=&-\frac{1}{2\pi}i e^{-(\Delta + \varepsilon_{p}) t} {\rm Ei}[(\Delta + \varepsilon_{p} - i\omega) t]\bigg|_{\omega},
\end{aligned}
\end{equation}
where ${\rm Ei}(z)=-\int^{\infty}_{-z}e^{-t}/tdt$ is the exponential integral function.
In the following, we use $g_{\psi\phi}=g_{b}$ throughout the paper,
i.e., the bare polaronic coupling which is transferred momentum-independent but may be scale-dependent
in some specific materials.
Note that the part of boson field without any order parameter fluctuation
$\phi(0)$ with $q=\Omega=0$ is also finite for a bose system
(like the phonon system) unless at zero-temperature with $k=\nu=0$.
In the presence of mean-field shift of boson field,
the anomalous self-energy (off-diagonal) can be written in terms of the anomalous Green's function $G_{A}(k,\nu)$,
self-consistently, as
\begin{equation} 
\begin{aligned}
\phi(0)
=&-g_{b}\frac{1}{\beta}\sum_{k,\nu}G_{A}(k,\nu)\\
=&-g_{b}\frac{1}{\beta}\sum_{k,\nu}\int^{\beta}_{0}dt e^{i\nu t}\langle\mathcal{T}\psi(k,t)\psi(-k,0)\rangle\\
=&-g_{b}\frac{1}{\beta}\sum_{k,\nu}\frac{\phi(0)}{G^{-2}(k,\nu)+\phi^{2}(0)},
\end{aligned}
\end{equation}
where $G(k,\nu)$ is the normal fermion Green's function dressed by normal self-energy.
This can be rewritten at low-temperature limit as
\begin{equation} 
\begin{aligned}
\phi(0)=-g_{b}
\frac{\phi(0)}{(i\nu-\varepsilon_{k})^{2}+\phi^{2}(0)}.
\end{aligned}
\end{equation}
This is a saddle-point equation,
and the finite-$q$ states induce the charge density fluctuation around its saddle-point solution.
Here the saddle-point solution about the anomalous self-energy is hard to directly obtained, but 
the mean-field saddle-point can be obtained
through the particle-hole correlation at mean field with $q=\Omega=0$ in the following.
$\phi(0)$ is also related to the gap equation through
(here we omit the subscript of $k,\nu$ in the $\phi(0)$ and $\phi(q)$)
\begin{equation} 
\begin{aligned}
\Delta=\int_{k,\nu}Z\phi(0)
+\int_{k,\nu}(1-Z)\phi(q),
\end{aligned}
\end{equation}
 where 
\begin{equation} 
\begin{aligned}
Z=&\frac{1}{1-g_{b}^{2}\partial\uppi(q,\Omega)/\partial \Omega\bigg|_{q=0}}\\
\approx & 
\frac{1}{1-g_{b}^{2}\uppi(i\Omega)/i\Omega)}
\end{aligned}
\end{equation}
 is the quasiparticle residue as a function of frequency in lowest order approximation
describing the weigth of nonfluctuating field-mediated pairing interaction.
The approximation in the last line is valid only in lowest order expansion in noninteracting case,
i.e., setting $q=0$.
Then there are stable and undamped bosons
and correspond to the propagating component of order parameter
resulting from the large particle-hole asymmetry.
The residue $Z=1$ corresponds to the absence of fluctuating boson field.
Here $g^{2}_{b}\uppi(i\Omega)$ is the bosonic self-energy (see Eq.(30)).
The emergence of nonfluctuating field $\phi(q=0)$ generally requires lower temperature
compares to that required by $\phi(q)$,
and it introduces the dissipationless nonresonance mode to the total order parameter.
Also, the undamped mode usually corresponds to the stronger 
coupling $g_{b}$ within the order parameter of boson field 
Other pairing correlation effects (like the Coulomb pseudopotential or Cooper instability
which are not included here)
can also be revealed by the off-diagonal anomalous self-energy.

The bosonic field in this paper is the particle-hole bubble type,
while for the case that the each propagator of
particle-hole bubble is the one-loop dressed Cooper pair\cite{Strack P}),
the bosonic mass term could appears within the action as $-g_{\psi\phi}^{-1}$.
The $\phi^{2}$-term related action reads
\begin{equation} 
\begin{aligned}
S_{\phi}
=&\frac{1}{2}\int\frac{d^{3}q}{(2\pi)^{3}}\phi^{*}(q)D^{-1}(q,\Omega)\phi(q),
\end{aligned}
\end{equation} 
where $D(q,\Omega)$ is the free order-parameter boson propagator
\begin{equation} 
\begin{aligned}
D(q,\Omega)
=&\frac{1}{-i\Omega+\Omega^{2}+q^{2}+m_{\phi}^{2}}.
\end{aligned}
\end{equation}
where
$m_{\phi}=g_{b}\phi_{q}$
is the order parameter mass term,
i.e., the order parameter with independent space and time.
Since the boson field with high energy scalar will strongly decayed into the quasiparticles
and thus leads to non-fermi liquid theory,
this free order-parameter propagator is valid only at small coupling and $\phi$, i.e., with vanishing gap term
(share the similar properties of massless particle and hole propagators) at low-enough temperature.
It is for sure that, the boson propagator with finite-$q$ must contains a decaying term when above the critical temperature,
to correctly describes the collective dynamics.
The boson field $\phi$ with low energy scalar means both the particle and hole (within the bubble)
are close to the fermi surface.

In imaginary time domain, it reads
\begin{equation} 
\begin{aligned}
D(q,\tau)
=&\int\frac{d\Omega}{2\pi}
\frac{e^{-i\Omega\tau}}{-i\Omega+\Omega^{2}+q^{2}+m^{2}_{\phi}}\\
=&
\frac{1}{2\pi\sqrt{C_{1}}}
(e^{\frac{1}{2} (1 - i \sqrt{C_{1}}) t}
    {\rm Ei}[\frac{1}{2} i (i - 2 \Omega + \sqrt{C_{1}}) t] - 
  e^{\frac{1}{2} (1 + i \sqrt{C_{1}}) t}
    {\rm Ei}[-(\frac{1}{2})       i (-i + 2 \Omega + \sqrt{C_{1}}) t])\bigg|_{\Omega},
\end{aligned}
\end{equation}
where $C_{1}=-1 - 4 m_{\phi}^{2} - 4 q^{2}$.
The linear-in-bosonic frequency term $-i\Omega$ in denominator of bosonic propagator $D(q,\Omega)$
implies that the dynamical critical exponent is larger than unity (and thus violates the Lorentz invariance),
which is caused by the particle-hole asymmetric due 
to the finite chemical potential of (doped) majority component ($\mu_{m}>\Omega,q$).
While at particle-hole symmetry with  Nambu construction,
we have
\begin{equation} 
\begin{aligned}
D(q,\Omega)
=&\frac{1}{\Omega^{2}+q^{2}+m^{2}_{\phi}},\\
D(q,\tau)=&
\frac{1}{2\pi}
\frac{1}{2 \sqrt{C_{2}}}
(e^{-i \sqrt{C_{2}} t} ({\rm Ei}[i (-\Omega + \sqrt{C_{2}}) t] - 
   e^{2 i \sqrt{C_{2}} t}
     {\rm Ei}[-i (\Omega + \sqrt{C_{2}}) t]))
\bigg|_{\Omega},
\end{aligned}
\end{equation}
where $C_{2}=-m_{\phi}^{2} - q^{2}$.
The dynamical critical exponent $z$ larger than 1 is usually related to the coupling to an emergent
gauge field or the relevant perturbations.
At fermionic quantum critical point ($Z_{2}$ Gross-Neveu class) and with particle-hole symmetry,
the scalar boson mode propagator reduced to the form
$D_{0}(q,\Omega)=\langle \phi(q,\Omega)\phi(-q,-\Omega)\rangle
=1/(\Omega^{2}+q^{2})$ 
The boson order-parameter field related gap equation at zero temperature 
has the following relation in mean-field approximation (with $q=0$ state and without the particle-hole fluctuation),
 $\Delta\propto |\phi_{q}|\propto 
([\int^{\Lambda}_{q,\Omega}D(q,\Omega)]^{-1}-[\int^{\Lambda}_{q,\Omega}D_{0}(q,\Omega)]^{-1})$,
$\int_{q,\Omega}D(q,\Omega)$ has a solution corresponds to bipolaron state
only when
$[\int^{\Lambda}_{q,\Omega}D(q,\Omega)]^{-1}
<[\int^{\Lambda}_{q,\Omega}D_{0}(q,\Omega)]^{-1}$.
The relaxation time of quasiparticle $1/\tau$ in an particle-hole order is proportional to the 
single-particle gap $\Delta$.
The above $\phi$-propagator is different in form to the propagator of single bosonic excitation 
which reads
\begin{equation} 
\begin{aligned}
D_{b}(q,\Omega)
=&\frac{1}{i\Omega-\varepsilon_{q}-\Sigma_{B}}.
\end{aligned}
\end{equation}
$\Sigma_{B}=-g_{b}^{2}\uppi(q,\Omega)
                    =-g_{b}^{2}\overline{\psi}_{k+q}\psi_{k}\overline{\psi}_{k}\psi_{k+q}=\phi^{2}(q)$ is the dynamical bosonic self-energy,
We note that, the inverstigation coupling between boson field propagator and fermion excitation propagator,
fermion field propagator and bosonic excitation propagator,
boson field propagator and fermion field propagator in one loop as well as two loop level
can be found in Refs.\cite{Joshi D G,Yu J,Han S E}.
While in this article,
we focus only on the coupling between bosonic excitation propagator and fermion excitation propagator.
Usually, in renormalization group (RG) analysis,
we focus on the
 infinitesmall ultra-violet shell $\Lambda e^{-z\ell}<\sqrt{\omega^{2}+\varepsilon_{p}^{2}}<\Lambda$
with the rescaling quantities $\omega\rightarrow \omega e^{-z\ell},p\rightarrow p e^{-z\ell}$,
the  scaling dimension at tree level of fermion field is larger than the bosonic field,
i.e.,
${\rm dim}[\psi]=3/2,{\rm dim}[\phi]=1$ in three spatial dimensions, and for Yukawa coupling, its scaling dimension
can be related to the dynamical exponent as ${\rm dim}[g_{b}]={\rm dim}[\omega]/4=z/4$.
The mass $m_{\phi}$ has a spatial-dimension-independent dimension, ${\rm dim}[m_{\phi}]=1$.
Note that in the absence of $q$-state at critical surface, $m_{\phi}=0$, and ${\rm dim}[\phi]=3/2$.
The related RG analysis can be found in Refs.\cite{Yu J,Joshi D G,Nikoli? P}.

As long as the bosonic (order parameter) mass term $m_{\phi}=g_{b}\phi_{q}$,
fully gaps the noninteracting quasiparticle (polaron) spectrum as $\sqrt{\varepsilon_{p}+|m_{\phi}|^{2}}$,
we have  $\{H_{0},g_{b}\phi\}\neq 0$ (or $\{S_{0},S_{g}\}\neq 0$.
While when $\{H_{0},g_{b}\phi\}\neq 0$
i.e., there exists another part of contribution to the fermion mass,
$m=m_{0}+g_{b}\phi$, then the order parameter mass term becomes
$g_{b}^{2}\phi^{2}+m_{0}^{2}+2m_{0}g_{b}\phi$ which is finite even in the presence of 
zero bosonic vacuum expectation value $\langle \phi\rangle$.
The related results for this case is discussed in Ref.\cite{Yu J}.
In fermion-induced criticality $m_{\phi}=0$,
the action is time reversal symmetry (under $\phi\rightarrow -\phi$),
and the lattice symmetries are always preserved in the absence of gauge fields,
e.g., 
for rotational invariant (in momentum space) anomalous self-energy $\phi_{k}(0)=-\phi_{-k}(0)$,
the normal Green's function preserves the lattice symmetry $G(k)=G(-k)$.
Note that near quantum critical point,
the quantum Yuakwa coupling as well as the boson field become irrelevant,
which is similar to what happen in high spatial-dimension\cite{Georges A,Janssen L,Yin S}.

Note that the charge is classically conserved in scalar Yukawa theory,
and the Yukawa potential is well-defined in the limit of large UV cutoff $\Lambda\rightarrow\infty$,
which corresponds to weak coupling limit $g_{b}\rightarrow 0$.
In real space the Yukawa potential is weak at large distance,
i.e., $V_{{\rm eff}}(x)\rightarrow -\infty$ as $x\rightarrow 0$
which corresponds to $V_{{\rm eff}}\rightarrow -\infty$ as $g\rightarrow -\infty$ ($\phi(q)\neq 0$)
 in the momentum space.
While the noninteracting system of majority particle can be described by the following action
\begin{equation} 
\begin{aligned}
S_{0}=\int\frac{d^{3}k}{(2\pi)^{3}}
\overline{\psi}(k)(i\Omega-\varepsilon_{k})\psi(k),
\end{aligned}
\end{equation}
similarly we can write the polaron's bare action as
\begin{equation} 
\begin{aligned}
S^{\downarrow}_{0}=\sum_{i=1,2}\int\frac{d^{3}p}{(2\pi)^{3}}
\overline{\psi}(p_{i})(i\omega_{i}-\varepsilon_{i})\psi(p_{i}).
\end{aligned}
\end{equation}
The order-parameter mass term $g_{\psi\phi}\phi$ does not contained in the quasiparticle energy term in bare actions.
While for the actions $S_{g}$ (or that of the polaron $S_{g}^{\downarrow}$),
$g_{\psi\phi}\phi=0$ can be treated as the quantum critical fixed point
(for $S_{g}^{\downarrow}$, the attractive bipolaron state disappear once $g_{\psi\phi}\phi\le 0$).
In the following text, we denote $\int_{k}=\int\frac{d^{3}k}{(2\pi)^{3}}$,
and unless otherwise specified, we work on the three spatial dimensions,
and the volume is setted as 1 for simplicity.

\section{Correlations and self-energies in strong-coupling and weak-coupling regime}

Firstly we consider the case of bipolaron that
the effective interaction is not one between two bare impurities
but the one between two impurities and the majority particles (fermi bath).
In this case the Yukawa interaction is results from the expansion of $T$-matrix 
to second order of $g_{b}$, similar to Frohlich-type contribution,
and is accurate only for weak coupling 
which related to the stable local minimum of Yukawa potential at vanishing $\psi=\phi(q)=0$.
Then in the absence of bosonic propagator,
the second order effective Yukawa potential
of a bipolaron system with weak attractive interaction,
can be written as
\begin{equation} 
\begin{aligned}
V_{{\rm eff}}=&-
   g_{b}^{2}\uppi(q,\Omega),
\end{aligned}
\end{equation}
where $\uppi(q,\Omega)$ is the particle-hole bubble which describes the density fluctuation.

The polaron-polaron coupling is nonlocal as long as the terms of $T$-matrix $T(p,\omega)$ 
with higher order of $g_{b}$ is considered.
That is to say,
the local contribution of bipolaron potential reads 
$V_{l}=-g_{b}^{2}\Pi(q,\Omega)$,
while the nonlocal contribution (exhibits many-body properties) of bipolaron potential reads
$V_{nl}=-T(p,\omega)^{2}\Pi(q,\Omega)$ where $T(p,\omega)$ is the $T$-matrix containing higher order terms 
of $g_{b}$.
Through numerical calculation we can know that
at weak coupling with $g_{b}\rightarrow 0$,
we have $g_{b}=T(p,\omega)$, i.e.,
the higher order terms of $g_{b}$ makes little effects and the local and nonlocal contributions
are almost the same;
while at strong coupling condition with large $|g_{b}|$,
we have $|g_{b}|\gg |T(p,\omega)|$,
i.e., the local approximation of the potential
overestimates the real one,
and in this case we must use the nonlocal contribution of bipolaron potential instead of the local one.
Thus the Yukawa potential can also be used in the presence of stronger coupling
as we replace the bare coupling by the $T$-matrix.
However, note that the Yukawa coupling mediated by bosons need to be strong enough
to make it possible to form the bipolaron,
i.e., the value of bipolaron self-energy $|{\rm Re}\Sigma_{bp}|$
need to larger than the ground state energy which can be approximately treated as chemical potential 
(of majority particles) at low-temperature limit,
since the binding energy reads $E_{bb}={\rm Re}\Sigma_{bp}+\mu_{m}$
and the formation of bipolaron requires $E_{bb}<0$
(note that we assume it is away from the classical limit and thus $\mu_{m}>0$).

We note that,
the pairing propagator during polaronic scattering is unlike the
one with two Matsubara Green's function merged into one particle-hole bubble,
i.e., the single-particle contribution which does not contributes to the effective pairing
interaction.
The vertices, which appears within vertex corrections
(like the case with spin-flip correlation),
is only required by the particle-hole channel scattering but not the Cooper channel one.

Then as shown diagrammatically in Fig.1(a), 
the coupling between two polarons can be described by the
$T$-matrices connected by a particle-hole loop,
and the whole diagram then contains two types of channels,
i.e., the particle-particle one (Cooper channel)
and particle-hole one. 
One may naively think the self-energy of such bipolaron configration is consist of the self-energies of the two single polarons
and the boson self-energy (the particle-hole loop),
i.e.,
\begin{equation} 
\begin{aligned}
\Sigma(p_{1},p_{2},\omega_{1},\omega_{2})=
\Sigma(p_{1},\omega_{1})+g_{b}^{2}\uppi(q,\Omega)+\Sigma(p_{2},\omega_{2}).
\end{aligned}
\end{equation}
However, this is incorrect in most cases.
From the perspective of ladder expansion to arbitary order of coupling $g_{b}$,
the self-energy of bipolaron should be
\begin{equation} 
\begin{aligned}
\Sigma(p_{1},p_{2},\omega_{1},\omega_{2})
=T_{1}(p_{1},\omega_{1})\uppi(q,\Omega)T_{2}(p_{2},\omega_{2}),
\end{aligned}
\end{equation}
where the non-self-consistent $T_{1}$-matrix reads
\begin{equation} 
\begin{aligned}
T(p_{1},\omega)
=\frac{1}{g_{b}^{-1}-\Pi(p_{1},\omega)}.
\end{aligned}
\end{equation}
Here the $\Pi(p,\omega)$ is the pair propagtor
\begin{equation} 
\begin{aligned}
\Pi(p_{1},\omega;k,\nu)
=&-   \int^{\beta}_{0}\int^{\beta}_{0}d\tau d\tau'
   e^{i\omega(\tau-\tau')}
   \langle \mathcal{T}
   (\sum_{q}\psi_{\uparrow}(p_{1}-q,\tau)\psi_{\downarrow}(k+q,\tau))
   (\sum_{q}\psi^{\dag}_{\uparrow}(p_{1}-q,\tau')\psi^{\dag}_{\downarrow}(k+q,\tau'))
   \rangle \\
=&\int_{q}\int\frac{d\Omega}{2\pi}
G_{\uparrow}(k+q,\nu+\omega)G_{\downarrow}(p_{1}-q,\omega-\Omega)\\
=&\int\frac{d^{3}q}{(2\pi)^{3}}
\frac{1-N_{F}(\varepsilon_{p_{1}-q})-N_{F}(\varepsilon_{k+q})}{\omega+i\eta-\varepsilon_{p_{1}-q}-\varepsilon_{k+q}}.
\end{aligned}
\end{equation}
By defining the Cooper-channel operator $F(\tau)=\sum_{q}\psi_{\uparrow}(p_{1}-q,\tau)\psi_{\downarrow}(k+q,\tau)$,
we have the relation (after analytical continuation)
 ${\rm Im}\Pi(Q,\mathcal{W})=-{\rm Im}\mathcal{W}^{-2}[\mathcal{F}(\mathcal{W})-\mathcal{F}(0)]$ in center-of-mass framework,
where $\mathcal{F}(\mathcal{W})=\int^{\beta}_{0}d\tau e^{i\omega\tau}\langle \mathcal{T}F(\tau)F^{\dag}(0)\rangle$.
Note that the Green's functions in second line of the above equation are bare propagators (undressed/unrenormalized by the self-energies),
and thus it is possible to observe the instability (satisfy the Thouless criterion at zero-center-of-mass momentum and at critical temperature),
as well as the gap equation\cite{?opik B}.
Through this pair propagtor, the pole of above non-self-consistent $T$-matrix, at zero center-of-mass momentum 
$Q=p_{1}+k=0,\mathcal{W}=\omega+\Omega=0$, gives rise to a divergence which leads to the instability and gap.
We can then rewrite the above pair propagator as
\begin{equation} 
\begin{aligned}
\Pi(Q,\mathcal{W})
=&\int_{q}
\frac{1-N_{F}(\varepsilon_{q+Q/2})-N_{F}(\varepsilon_{-q+Q/2})}{\mathcal{W}+i\eta+i\eta'-\varepsilon_{-q+Q/2}-\varepsilon_{q+Q/2}}.
\end{aligned}
\end{equation}
Then the pole at $Q=\mathcal{W}=0$ corresponds to
\begin{equation} 
\begin{aligned}
g_{b}^{-1}=\int_{q}
[
\frac{1}{2 \mu_{m} - 2\varepsilon_{q}}
  -\frac{2}{(2 \mu_{m} - 2\varepsilon_{q}) (1 + {\rm Cosh}[\mu_{m}\beta - \varepsilon_{q}\beta] - 
    {\rm Sinh}[\mu_{m}\beta - \varepsilon_{q}\beta])}
],
\end{aligned}
\end{equation}
or 
\begin{equation} 
\begin{aligned}
1=g_{b}\int_{q}
[
\frac{1}{2 \mu_{m} - 2\varepsilon_{q}}
  -\frac{2}{(2 \mu_{m} - 2\varepsilon_{q}) (1 + {\rm Cosh}[\mu_{m}\beta - \varepsilon_{q}\beta] - 
    {\rm Sinh}[\mu_{m}\beta - \varepsilon_{q}\beta])}
],
\end{aligned}
\end{equation}
where $\varepsilon_{q}=q^{2}/2m$.
For case that the polaron spectrum is gapped by the gap equation,
we have
\begin{equation} 
\begin{aligned}
1=&g_{b}\int_{q}
[
\frac{1}{2 \mu_{m} - \varepsilon_{q} - \sqrt{\Delta^{2} + \varepsilon^{2}_{q}}}\\
&   - \frac{1}{(2 \mu_{m} - \varepsilon_{q} - \sqrt{
    \Delta^{2} + \varepsilon^{2}_{q}}) (1 + {\rm Cosh}[\mu_{m}\beta - \varepsilon_{q}\beta] - 
    {\rm Sinh}[\mu_{m}\beta - \varepsilon_{q}\beta])}\\
&	- \frac{1}{(2 \mu_{m} - \varepsilon_{q} - 
	\sqrt{\Delta^{2} + \varepsilon^{2}_{q}}) (1 + 
    {\rm Cosh}[\mu_{m}\beta - \sqrt{\Delta^{2} + \varepsilon^{2}_{q}}\beta] - 
    {\rm Sinh}[\mu_{m}\beta - \sqrt{\Delta^{2} + \varepsilon^{2}_{q}}\beta])}
].
\end{aligned}
\end{equation}
The self-consistent solution is available at low-temperature where $Z\sim 1$
(but note that $Z\neq 1$ even at zero temperature unless for phonons, since we assume that there are finite-$q$ state even at zero temperature,
which is similar to the form of Pippard kernel).
Note that the self-consistent solution cannot be obtained at temperatures higher than the critical one\cite{Ohashi Y2,Beach K S D} 
where $1>g_{b}\Pi(0,0)$ and thus there is no instabilities, unless in thermodynamical limit.
Obviously, the pair propagator is vanishingly small as $\Delta\rightarrow\infty$
(then renormalization to $T$-matrix is absent $T\rightarrow g_{b}$),
and thus requires $g_{b}\rightarrow\infty$.
That may leads to the instability as well as the non-self-consistent solution according to Thouless criterion.
The Thouless criterion can also be written in a Ward-identity-like form,
$T^{-1}(Q=0,\mathcal{W}=0)\Delta=0$,
i.e., the $T$-matrix is divergent as long as the order-parameter is nonzero,
which corresponds to the zero-energy collective mode.
Furthermore, since at $q=0$
the self-energy of bipolaron is zero (as can be easily seen from Eq.()),
i.e., the Ward-identity $T=\partial \Sigma/\partial (i\omega)$ is broken,
and
the nonfluctuating order parameter $\phi(0)$ cannot affects this self-energy,
i.e., the change of self-energy in phase space has $\delta\Sigma(p,\omega)=\delta (H-H_{0})/\delta G(p,\omega)=0$
(here $H-H_{0}\sim g_{b}/N$; $N$ is teh flavor number),
thus the gap equation within above formula can simply be replaced by the fluctuating order parameter $\phi(q)$.
Using the Ward identity\cite{Haussmann R}, which is valid in both the self-consistent and non-self-consistent approximation\cite{?opik B},
the order parameter here can also be replaced by the variation of the self energy 
$\delta\Sigma(p,\omega)=\sum_{q,\Omega}T(q,\Omega)\delta G(p-q,\omega-\Omega)$.
Thus the poles of $T$-matrix also leads to the divergence of self-energy.

The divergent part of non-self-consistent $T$-matrix,
in Eliashberg-Migdal approximation (i.e., to first-order of $g_{b}$) is seperable, then
the contribution of vanishing mode $(Q=0,\mathcal{W}=0)$ to the bipolaron self-energy 
reads $\Sigma=-\phi^{*}(q)\phi(q)$, and it vanishes at thermodynamic limit,
where we donot consider the contribution of exchange digrams to the self-energy.
The divergence of the $T$-matrix (particle-particle pairing susceptibility) at critical temperature
indicates the instability, which is similar to the case happen in superconductivity transition\cite{Dong X,Macridin A}.
For the case of $Z\rightarrow 0$, and considering an infinite number of bosonic bubbles,
we have the zero mode contribution
\begin{equation} 
\begin{aligned}
\Sigma=-\phi_{k}^{*}(q)\phi_{k}(q)=-(-\frac{1}{\beta}\sum_{k'}g_{b}G_{k+q}G_{k}\phi_{k'}(q))^{*}
(-\frac{1}{\beta}\sum_{k'}g_{b}G_{k+q}G_{k}\phi_{k'}(q)).
\end{aligned}
\end{equation}
This self-energy (in first order approximation of coupling) is shown in diagram Fig.,
but note that in infrared limit with vanishing $q$, the higher-loop corrections are suppressed by a large number of fermion flavor.
For vanishing $q$, the mean field contribution to anomalous self-energy ($\Sigma_{MF}=\Delta$) is shown diagrammatically in Fig.(b),
whose action about the side-interaction reads 
\begin{equation} 
\begin{aligned}
S_{SI}=g\int_{p}\psi_{p}[\int_{p'}G(p')]\psi^{\dag}_{p}.
\end{aligned}
\end{equation}
There also exists another type of bosonic diagram as shown in Fig.(c) which has an infinite number of bubbles
but different from the one shown in Fig.(a).
The boson propagator of diagram in Fig.(c) reads (in imaginary time domain)
\begin{equation} 
\begin{aligned}
D_{0}(\tau)
=&-\langle \mathcal{T}D_{1}(\tau)D_{2}(\tau_{1})\cdot\cdot\cdot D_{n}(0)\rangle\\
=&-e^{\tau(\varepsilon_{k_{1}+q}-\varepsilon_{k_{1}})}
[\delta_{q,0}\prod_{i=1}^{n}N_{F}(\varepsilon_{k_{i}})
+\delta_{k_{1}+q,k_{i;i\neq 1}}N_{F}(\varepsilon_{k_{1}+q})(1-N_{F}(\varepsilon_{k_{1}}))
].
\end{aligned}
\end{equation}

The vanishing limit of $q$ also corresponds to the strong coupling limit
(see Eq.()),
in which case the pair propagator behaves like a gapless boson in broken-symmetry phase\cite{Pieri P,Lee S S}.
While for the criticality at zero-temperature limit (also corresponds to the long-wavelength limit $q\rightarrow 0$),
\begin{equation} 
\begin{aligned}
\Sigma=-\phi^{*}(0)\phi(0)=-g_{b}^{2}\frac{\phi^{2}(0)}{[(i\nu-\varepsilon_{k})^{2}+\phi^{2}(0)]^{2}}.
\end{aligned}
\end{equation}
Furthermore, at zero-temperature limit the total bosonic momentum is proportional to the conserved polaron momentum\cite{Dzsotjan D},
and thus vanishes at $Q=0$.
Here the term $\frac{\phi^{2}(0)}{[(i\nu-\varepsilon_{k})^{2}+\phi^{2}(0)]^{2}}$
can be viewed as a modified self-consistent propagator between two impurities in dynamical mean field theory,
and the above results reflect that,
the bipolaron self-energy becomes independent of the initial impurities momenta ($p_{1}$ and $p_{2}$)
in Eliashberg-Migdal approximation, where the gap equation can also be viewed as a modified Eliashberg equation in this case,
and there are only off-diagonal elements (the mass term) exist in the self-energy matrix 
($g_{b}\rightarrow 0$ for $q\rightarrow 0$ or $\Lambda\rightarrow 0$).

$\uppi(q,\nu)$ is the noninteracting density-density correlation (response) function
\begin{equation} 
\begin{aligned}
\uppi(q,\Omega)
=&-\int^{\beta}_{0}d\tau e^{i\Omega \tau}
\langle\mathcal{T}\rho(q,\tau)\rho(-q,0)\rangle\\
=&\frac{1}{\beta}\sum_{\nu}
\int_{k}
G(k,\nu)G(k+q,\nu+\Omega)\\
=&\int_{k}
\frac{N_{F}(s\varepsilon_{k})-N_{F}(s'\varepsilon_{k+q})}{\Omega+i\eta'+s\varepsilon_{k}-s'\varepsilon_{k+q}}.
\end{aligned}
\end{equation}
Note that the Pauli matrices should also be included in the expression
when the amplitude or phase fluctuations are also considered 
(like the spin/peudospin-orbit coupling) but not just the density fluctuation.
The above density-density correlation formula is well-defined for small transferred momentum $q$,
i.e., the long-range correlation,
and the density fluctuation here also reflects the fluctuation of order parameter,
especially at strong coupling.
The above formula can be rewritten as
\begin{equation} 
\begin{aligned}
\uppi(q,\Omega)
=&\int_{k}
[
\frac{N_{F}(-\varepsilon_{k})-N_{F}(\varepsilon_{k+q})}{\Omega+i\eta'-\varepsilon_{k}-\varepsilon_{k+q}}
+\frac{N_{F}(\varepsilon_{k})-N_{F}(-\varepsilon_{k+q})}{\Omega+i\eta'+\varepsilon_{k}+\varepsilon_{k+q}}\\
&
+\frac{N_{F}(-\varepsilon_{k})-N_{F}(-\varepsilon_{k+q})}{\Omega+i\eta'-\varepsilon_{k}+\varepsilon_{k+q}}
+\frac{N_{F}(\varepsilon_{k})-N_{F}(\varepsilon_{k+q})}{\Omega+i\eta'+\varepsilon_{k}-\varepsilon_{k+q}}
],
\end{aligned}
\end{equation}
where the first and second terms describe the interband electron-hole transition
that leads to damping at energy above gap $\Omega>2\Delta$,
 while the third and fourth terms
describe the intraband electron-hole transition that
leads to the Landau damping (at finite temperature)
with energy below gap $\Omega<2\Delta$.
So the interband part becomes important as the gap becomes smaller.
We assume that $\varepsilon_{k+q}>\varepsilon_{k}>\mu_{m}$
and $\varepsilon_{-(k+q)}<\varepsilon_{-k}<\mu_{m}$),
then at zero-temperature the intraband part is absent,
we obtain
\begin{equation} 
\begin{aligned}
\uppi(q,\Omega)
=\int_{k}
\frac{2(\varepsilon_{k}+\varepsilon_{k+q})}
{(\Omega+i\eta')^{2}-(\varepsilon_{k}+\varepsilon_{k+q})^{2}}.
\end{aligned}
\end{equation}
This is in consistent with the intrinsic case (undoped).
While for intraband transition,
the particle-hole polarization becomes
\begin{equation} 
\begin{aligned}
\uppi(q,\Omega)
=\int_{k}
\frac{2(\varepsilon_{k}-\varepsilon_{k+q})}
{(\Omega+i\eta')^{2}-(\varepsilon_{k}-\varepsilon_{k+q})^{2}}.
\end{aligned}
\end{equation}
For interband electron-hole transition, We find that
\begin{equation} 
\begin{aligned}
\uppi(0,0)
=&
-16 m \pi (\frac{k}{2} + 
   \frac{1}{8} (-\frac{m^{2} z^{2} {\rm ln}[k - \sqrt{-i m z}]}{(-i m z)^{3/2}} + \frac{
      m^{2} z^{2} {\rm ln}[k + \sqrt{-i m z}]}{(-i m z)^{3/2}} \\
&- \frac{
      m^{2} z^{2} {\rm ln}[k - \sqrt{i m z}]}{(i m z)^{3/2}} + \frac{
      m^{2} z^{2} {\rm ln}[k + \sqrt{i m z}]}{(i m z)^{3/2}}))\bigg|_{k}.
\end{aligned}
\end{equation}
Since at $q=0$, $Z=1$, then
the gap equation has
\begin{equation} 
\begin{aligned}
\Delta
=&\int_{k,\nu}Z\phi(0)\\
=&\int_{k,\nu}\phi(0)\\
=&\int_{k,\nu}g^{2}_{b}\uppi(0,0)\\
\end{aligned}
\end{equation}
i.e., the mean-field saddle-point solution of anormal self-energy can be simply written as
$\phi_{k}(0)=g^{2}_{b}\uppi(0,0)$, since the anomalous self-energy always follows from the structure of $T$-matrix.
When infinite bubbles are included, we can also use the form
$\phi_{k}(0)=g_{b}\uppi(0,0)\phi_{k'}(0)$.

The final expression of the diagonal element of self-energy matrix of bipolaron mode reads ($q_{1}<q$)
\begin{equation} 
\begin{aligned}
&\Sigma(p_{1},p_{2},\omega_{1},\omega_{2})\\
&=\int_{q}
[
\frac{1}{g_{b}^{-1}-\Pi(p_{1},\omega_{1})}
\uppi(q,\Omega)
\frac{1}{g_{b}^{-1}-\Pi(p_{2},\omega_{2})}
]\\
&=\int_{q}
[
\frac{1}{g_{b}^{-1}-\int_{q_{1}}G(p_{1}-q_{1},)G(k+q_{1})}
\int_{k}G(k+q,\nu+\Omega)G(k,\nu)
\frac{1}{g_{b}^{-1}-\int_{q_{1}}G(p_{2}+q_{1})G(k+q-q_{1})}
]\\
&=\int_{q}
[
\frac{1}{g_{b}^{-1}-\int_{q_{1}}G(p_{1}-q_{1},)G(k+q_{1})}
\frac{|\Delta|^{2}}{g_{b}^{2}}
\frac{1}{g_{b}^{-1}-\int_{q_{1}}G(p_{2}+q_{1})G(k+q-q_{1})}
],
\end{aligned}
\end{equation}
which is deduced in detail in Appendix.A.
Since we discuss in spatial dimension $d=3$, it is obviously than this diagonal element of self-energy matrix 
vanishes at $q=q_{1}=0$,
i.e., then the mean field self-energy of bipolaron reads
$\Sigma=\begin{pmatrix}
0 & \phi(0)\\
\phi^{*}(0)  & 0
\end{pmatrix}$.
For convenience of numerical study,
we still use the center-of-mass framework.
The pair propagator of first polaron (with impurity momentum $p_{1}$)
is shown in Eq.(22),
then the pair-propagator of second polaron reads
\begin{equation} 
\begin{aligned}
\Pi(Q',\mathcal{W'})
=&\int_{q}
\frac{1-N_{F}(\varepsilon_{q_{1}+Q'/2})-N_{F}(\varepsilon_{-q_{1}+Q'/2})}{\mathcal{W}+i\eta+i\eta'-\varepsilon_{-q_{1}+Q'/2}-\varepsilon_{q_{1}+Q'/2}},
\end{aligned}
\end{equation}
where $Q'=k+q+p_{2}=Q+2q$, $\mathcal{W'}=2\nu+2\Omega$.
At critical temperature (for the formation of polaron), 
since the self-energy (and the vertex function $T$-matrix) with zero center-of-mass momentum $Q=0$ is dominating,
we set $Q=0(Q'=2q)$ and $\Omega=0$,
then we can write
\begin{equation} 
\begin{aligned}
1-N_{F}(\varepsilon_{-q_{1}+Q/2})-N_{F}(\varepsilon_{q_{1}+Q/2})
=&1-N_{F}(\varepsilon_{q_{1}+Q/2})\\
=&1 - 1/(1 + {\rm cosh}[\beta \frac{q_{1}^2}{2 m}] + {\rm sinh}[\beta \frac{q_{1}^2}{2 m}])\\
=&\frac{1}{2}+\frac{\beta q_{1}^{2}}{8m}+O(q_{1}^{3}),\\
1-N_{F}(\varepsilon_{-q_{1}+Q'/2})-N_{F}(\varepsilon_{q_{1}+Q'/2})
=&1-N_{F}(\varepsilon_{-q_{1}+Q'/2})\\
=&1 - 1/(1 + {\rm cosh}[\beta \frac{(q-q_{1})^2}{2 m}] + {\rm sinh}[\beta \frac{(q-q_{1})^2}{2 m}]),\\
N_{F}(-\varepsilon_{k})-N_{F}(\varepsilon_{k+q})
=&\frac{1}{1 + {\rm cosh}[\beta\frac{k^{2}}{2 m}] - {\rm sinh}[\beta\frac{k^{2}}{2 m}]} - \frac{1}{
 1 + {\rm cosh}[\beta\frac{(k+q)^{2}}{2 m}] + {\rm sinh}[\beta\frac{(k+q)^{2}}{2 m}]}\\
\approx &\frac{1}{2} {\rm tanh}[\beta\frac{q^{2}}{4 m} ],
\end{aligned}
\end{equation}
for the first polaron, second polaron, and fermionic  bubble, respectively,
where we see that the formation of bipolaron at finite temperature requires larger value of $q$.
Then the two $T$-matrices can be obtained by inserting Eqs.(41,42) into Eq.(26).

While at zero temperature, similarly,
the two $T$-matrices can be obtained as
\begin{equation} 
\begin{aligned}
T_{1}(\omega_{1})=&[g_{b}^{-1}-
4 \pi (-q_{1} m + 
   m^{3/2} \sqrt{\mathcal{W} + i \eta} {\rm atanh}\frac{q_{1}}{\sqrt{m} \sqrt{\mathcal{W} + i \eta}})\bigg|_{q_{1}}]^{-1},\\
T_{2}(\omega_{2})=&[g_{b}^{-1}+
   4 m \pi (q_{1} - 
   \sqrt{q^{2} - m (\mathcal{W} + i \eta)} {\rm atan}\frac{q_{1}}{\sqrt{q^{2} - m (\mathcal{W} + i \eta)}})\bigg|_{q_{1}}]^{-1}.
\end{aligned}
\end{equation}
While  the boson self-energy reads
\begin{equation} 
\begin{aligned}
(2\pi)^{3}\uppi(q,\Omega)=&
-
   m \pi (4 k + (
   8 m (\Omega + i \eta) {\rm atan}[\frac{2 k + q}{\sqrt{
     q^{2} - 4 m (\Omega + i \eta)}}])/\sqrt{q^{2} - 4 m (\Omega + i \eta)} \\
	 &+ 
   2 i q {\rm atan}[\frac{2 m \eta}{2 k^{2} - 2 m \Omega + 2 k q + q^{2}}] \\
   &- 
   q {\rm ln}[4 k^{4} + 8 k^{3} q - 4 m \Omega q^{2} + q^{4} + 
      k^{2} (-8 m \Omega + 8 q^{2}) + k (-8 m \Omega q + 4 q^{3}) + 
      4 m^{2} (\Omega^{2} + \eta^{2})])\bigg|_{k},
\end{aligned}
\end{equation}
or at zero $\Omega$, it simplfied as
\begin{equation} 
\begin{aligned}
(2\pi)^{3}\uppi(q,0)=&
-
   8 m \pi (\frac{k}{2} + \frac{1}{4} i q {\rm atan}[(2 m \eta)/(2 k^{2} + 2 k q + q^{2})] + \frac{
   i m \eta {\rm atan}[\frac{2 k + q}{\sqrt{q^{2} - 4 i m \eta}}]}{\sqrt{
   q^{2} - 4 i m \eta}} \\
&- 
   \frac{1}{8} q {\rm ln}[
     4 k^{4} + 8 k^{3} q + 8 k^{2} q^{2} + 4 k q^{3} + q^{4} + 4 m^{2} \eta^2])\bigg|_{k}.
\end{aligned}
\end{equation}
Note that during all the numerical simulations,
we set the UV cutoff as $\Lambda_{k}=\Lambda_{q}=2\Lambda_{q_{1}}$.
And the attractive polaronic coupling within each single polaron is setted as $g_{b}=-1$.
The results are shown in Fig..
During caluclating the self-energies, we set $\Omega=0$
(in which case $Z\approx 0$ and $\Delta\approx \int_{k,\nu}\phi(q)$),
and thus the simpler expression Eq.(44) is applied and we have $\mathcal{W}=\mathcal{W'}$.
But During calculating the quasiparticle residue (spectral weight of the quasiparticle peak) $Z(\Omega)$,
we must keep $\Omega$ finite but setting $q=0$ as we explained in above.
As revealed in the Fig.4, both the two single-polarons exhibit attractive feature (negative self-energy)
at large frequency $\mathcal{W}$ where the imaginary self-energy turns to the value of $g_{b}$:
${\rm \Sigma}\rightarrow g_{b}$, and at higher temperature, this turning process will becomes slower.
We can also see that, the bipolaron formed by two attractive fermion polarons also exhibits the attractive feature.
From the imaginary part of bipolaron self-energy,
we see that $|{\rm Im}\Sigma|\sim\mathcal{W}^{\alpha}$ with $\alpha>1$ here
(note that the value of $\alpha$ is indeed affected by the spatial dimension),
and thus the bipolaron lifetime has $\tau\sim\mathcal{W}^{-\alpha}$.
Physically, this also related to the long lifetime in diffusion channel (particle-hole) especially at weak-coupling regime with large-$q$ states
(see Fig.).
While for the zero temperature bosonic self-energy $\uppi(q)$ as shown in Fig.,
we found that $\uppi(q)\sim |q|$ near the critical transition point $q=0$,
which is in consistent with the bosonic self-energy evaluated in one-loop diagram
(without consider the vertex corrections) in Fermi-liquid state.
By comparing the bosonic self-energies at zero-temperature (Fig.5) and at finite-temperature (Fig.7),
we found that $|\uppi(q)|$ decreases with the increasing temperature,
and it vanishes at large $q$ as well as zero $q$ states.
At finite temperature, the bipolaron still has a negative self-energy (Fig.6)
at large frequency,
whose absolute value is, however, heavily decreased by the increasing temperature.
From Fig.6,
we found that,
for single-polaron, the absolute value of self-energy weakly increased with the increasing temperature at large frequency,
but for bipolaron, the absolute value of self-energy decreased with the increasing temperature at large frequency.
That reveals a change of
behavior when two polarons combine, through a boson excitation, into a bipolaron.
The quasiparticle residue is shown in Fig.8,
where we found that, at zero $\Omega$, $Z=0$,
and as the temperature increases, the noninteracting effect becomes dominating,
which is in agreement with Ref.\cite{Abrahams E}.

The fermi-liquid feature of bipolaron can be seen from the self-energy in the (third panel of) Fig.4 and Fig.6,
which behaves as $|{\rm Re}\Sigma|\sim \mathcal{W}^{2}$ in the low-frequency regime,
although this behavior turns gradually to be $|{\rm Re}\Sigma|\sim \mathcal{W}$ as the temperature increases.
Through the spectral functions shown in Fig. and Fig.,
the shape and symmetry peaks exhibit mainly the fermi-liquid character,
especially at zero- and low-temperature.
There have three main reasons for this: firstly,
the non-fermi-liquid can be caused by coupling the fermi surface to massless boson at critical point,
however, there exists a finite bosonic mass (gapped boson field) due to the strong screening effect origins from the particle-hole excitations;
secondly,
the interaction is dominated by the short-range one (also due to the screening effect);
thirdly,
the particle-hole propagator (RPA diagram) in Eq.(41) does not contain the logarithmic singularities as 
shown in Fig.(5) and Fig.(7).


\section{Aslamazov-Larkin-type diagram}

Next we explore the other contributions to the correlation between two impurities,
as shown by the Aslamazov-Larkin-type diagram in Fig. with external momentum $Q'-Q=2q$,
where the outer bubble formed by the impurity propagators in symmetry-broken phase provides the anorther fluctuation contribution
other that the inner particle-hole bubble.
Note that
unlike the case of single-polaron, where the contribution to conductivity is negative\cite{Galitski V M},
the contribution of Aslamazov-Larkin-type diagram here to conductivity of fluctuating polarons is positive.
Since the random-phase approximation (RPA), as a mean-field-level approximations,
fails to describes the charge-density-wave instability,
the Aslamazov-Larkin vertex correction is necessary in calculating the correlation function between two impurities.
\begin{equation} 
\begin{aligned}
\chi(q)=&\frac{1}{\beta}
\int_{Q}
T_{1}(Q)T_{2}(Q')
G_{i}(\frac{Q}{2}-q)G_{i}(\frac{Q}{2}+q)G_{m}(Q'/2)
G_{i}(\frac{Q}{2})G_{m}(\frac{Q'}{2}-q)G_{i}(Q'/2+q),
\end{aligned}
\end{equation}
where $G_{i}$ and $G_{m}$ denote the bare propagators  of impurity and majority particle.

In matrix form,
since 
\begin{equation} 
\begin{aligned}
{\bf T}(p)=-g_{b}\sigma_{0}-g_{b}{\bf \Pi}(p){\bf T}(p),
\end{aligned}
\end{equation}
we has the solution 
\begin{equation} 
\begin{aligned}
{\bf T}(p)=\frac{1}{(g_{b}^{-1}-\Pi_{22}(p))(g_{b}^{-1}-\Pi_{11}(p))-\Pi_{12}(p)^{2}}
\begin{pmatrix}
g_{b}^{-1}-\Pi_{11}(p) & \Pi_{12}(p)+\phi(0)\\
\Pi_{21}(p)+\phi(0)  & g_{b}^{-1}-\Pi_{22}(p)
\end{pmatrix}.
\end{aligned}
\end{equation}
where the diagonal elements in the matrix describe the fluctuations while the non-diagonal elements in the matrix describe
the coupling between fluctuations (the first term) and the mean-field gap term (the second term).
In strong-coupling limit with $\Lambda_{q}\rightarrow 0$,
we have
\begin{equation} 
\begin{aligned}
\Pi_{11}&=-\int_{q_{1}}\frac{1}{\beta}
\sum_{n}G_{11}(p_{1}-q_{1})G_{11}(k+q_{1})\\
&=-\int_{q_{1}}
\frac{m ({\rm coth}[\frac{q_{1}^{2} + 2 q_{1} k + k^{2} - 2 i m \nu}{4 m T}] + 
   {\rm coth}[\frac{q_{1}^{2} - 2 q_{1} p + p^{2} - 2 i m \omega}{4 m T}])}{
2 q_{1}^{2} + k^{2} - 2 i m \nu + 2 q_{1} (k - p) + p^{2} - 2 i m \omega}\\
&\approx
\int_{q_{1}}\frac{m}{q^{2}_{1}},
\end{aligned}
\end{equation}
where the Matsubara frequency of boson excitation is $\Omega_{n}=2n\pi T$.
and $G_{11}(p,\omega)=\frac{i\omega-\varepsilon_{p}}{(i\omega-\varepsilon_{p})^{2}+\Delta^{2}}
\approx \frac{1}{i\omega-\varepsilon_{p}}$ since $\Delta$ is vanishingly small at strong-coupling limit.
The approximation in the last line is valid at high temperature.
Note that here we apply the non-self-consistent approximation,
that the Green's functions are not dressed:
\begin{equation} 
\begin{aligned}
G_{11}(p,\omega)=&\frac{i\omega-\varepsilon_{p}}{(i\omega-\varepsilon_{p})^{2}+\Delta^{2}}=-G_{22}(-p,-\omega),\\
G_{12}(p,\omega)=&\frac{\Delta}{(i\omega-\varepsilon_{p})^{2}+\Delta^{2}}=G_{21}^{*}(p,\omega);
\end{aligned}
\end{equation}
while in self-consistent case,
the above equations should be
\begin{equation} 
\begin{aligned}
G_{11}(p,\omega)
=&\frac{i\omega+\varepsilon_{-p}-\Sigma_{22}(p,\omega)}
{(i\omega-\varepsilon_{p}-\Sigma_{11}(p,\omega))(i\omega+\varepsilon_{-p}-\Sigma_{22}(p,\omega))-\Delta^{2}}\\
=&\frac{-i\omega-\varepsilon_{-p}+\Sigma_{22}(p,\omega)}
{(i\omega-\varepsilon_{p}-\Sigma_{11}(p,\omega))(-i\omega-\varepsilon_{-p}+\Sigma_{22}(p,\omega))+\Delta^{2}},\\
G_{12}(p,\omega)
=&\frac{\Delta}
{(i\omega-\varepsilon_{p}-\Sigma_{11}(p,\omega))(i\omega+\varepsilon_{-p}-\Sigma_{22}(p,\omega))-\Delta^{2}}\\
=&\frac{\Delta}
{(i\omega-\varepsilon_{p}-\Sigma_{11}(p,\omega))(-i\omega-\varepsilon_{-p}+\Sigma_{22}(p,\omega))+\Delta^{2}},\\
\end{aligned}
\end{equation}
with the self-energies reads
\begin{equation} 
\begin{aligned}
\Sigma_{11}^{-1}(p,\omega)=&g_{b}^{-1}-\Pi_{11}(p,\omega),\\
\Sigma_{22}^{-1}(p,\omega)=&g_{b}^{-1}-\Pi_{22}(p,\omega)=\Sigma_{11}(-p,-\omega).
\end{aligned}
\end{equation}
And the sum rule of spectral weight is satisfied (analytical continued) as
\begin{equation} 
\begin{aligned}
&-\frac{1}{\pi}\int^{\infty}_{-\infty}d\omega {\rm Im}G_{11}(p,\omega)=0,\\
&-\frac{1}{\pi}\int^{\infty}_{-\infty}d\omega {\rm Im}G_{12}(p,\omega)=0,
\end{aligned}
\end{equation}
and becomes, in large $\omega$ limit,
\begin{equation} 
\begin{aligned}
&-\frac{1}{\pi}\int^{\infty}_{-\infty}d\omega {\rm Im}G_{11}(p,\omega)=1,\\
&-\frac{1}{\pi}\int^{\infty}_{-\infty}d\omega {\rm Im}G_{12}(p,\omega)=0.
\end{aligned}
\end{equation}
Note that the sum rule in self-consistent case requires $\Sigma_{11}\rightarrow 0$
when $|i\omega|\rightarrow\infty$ as can be easily seen from Eq.(52).
As we presented above, in strong coupling limit (with large $q$), the $\Pi_{11}$ and $\Pi_{22}$  as well as the mean field gap term $\phi(0)$
 are subleading with respect to the term $\Pi_{12}(\Pi_{21})$,
we omit the small constant $\phi(0)$ in the following.

Then we have, in this limit,
\begin{equation} 
\begin{aligned}
g_{b}^{-1}-\Pi_{11}(p_{1})
=&
-\int_{q_{1}}\frac{1}{E_{b}+\varepsilon_{p_{1}-q_{1}}+\varepsilon_{k+q_{1}}}+
\int_{q_{1}}\frac{1}{\beta}
\sum_{n=-\infty}^{\infty}G_{11}(p_{1}-q_{1})G_{11}(k+q_{1})\\
\approx &  \int_{q_{1}}\frac{1}{2\varepsilon_{q_{1}}}-\int_{q_{1}}\frac{m}{q^{2}_{1}}\\
=&0,
\end{aligned}
\end{equation}
and similarly $g_{b}^{-1}-\Pi_{22}(p)\approx 0$.
Here we approximate the bare coupling to the gap equation
by assuming the static self-energy of single polaron is small enough and thus the bottom of
bound state band is close to zero.
The binding energy $E_{b}\sim a_{F}^{-2}$ vanishes only in unitarity zero-coupling limit,
and we have $E_{b}\sim |\Sigma_{p=0}|\sim|-\Pi(p=0)^{-1}|$\cite{Combescot R,Sidler M}.
Note that the high-temperature strong-coupling here is different from the
case of low-temperature tightly bound limit (adiabatic) as studied in 
Refs.\cite{Yildirm T,Pieri P,Andrenacci N,Mulkerin B C},
while in the latter case,
the binding energy cannot be neglected and the above expression turns out to be $\sim(\sqrt{(mE_{b})^{-1}}-\sqrt{(m\eta)^{-1}})$ with $\eta\rightarrow 0$.
Thus, due to the cutoff-dependent coupling, the particle-particle loop contribution (in ladder approximation)
 to the correlation function vanishes in strong coupling limit.
Since the amplitude, phase, or density fluctuations are described by the diagonal terms of pair propagator function,
these fluctuations vanish in strong coupling limit, unless they are coupled with eachother.
But that is different to the RPA bubble whose contribution to correlation function also vanishes in strong coupling limit
unless when the vertex correction is contained\cite{Andrenacci N}.
For real gap equation,
 the anomalous Green's function has $G_{12}=G_{21}^{*}$ (in BCS-type approximation or in the limit of large complex plane radius $|i\omega|$
in which case $G_{11}(\omega)\sim (i\omega)^{-1}$ and $G_{12}(\omega)\sim \frac{\Delta}{ (i\omega)^{2}-\Delta^{2}}$),
thus we have $\Pi_{12}=\Pi_{21}$.
Then the resulting $T$-matrix can be obtained as
\begin{equation} 
\begin{aligned}
{\bf T}(p)=
\begin{pmatrix}
0 & \frac{-1}{\Pi_{12}(p)}\\
\frac{-1}{\Pi_{12}(p)}  & 0
\end{pmatrix}.
\end{aligned}
\end{equation}
The off-diagonal element of pair propagator reads
\begin{equation} 
\begin{aligned}
\Pi_{12}=&\int_{q_{1}}\frac{1}{\beta}
\sum_{n}G_{12}(p_{1}-q_{1})G_{12}(k+q_{1})\\
\approx & \int_{q_{1}}\frac{\Delta m^{2}}{q_{1}^{2}}.
\end{aligned}
\end{equation}
Note that for weak-coupling case,
the $\Delta$ in above expression can simply be replaced by $\phi(0)$, and
similar to Eq.(56),
the approximation in second line requires high enough temperature.

In the zero-center-of-mass framework ($(Q,\mathcal{W})=(0,0)$),
we can similarly obtain, for the second polaron,
\begin{equation} 
\begin{aligned}
{\bf T}_{2}(0)=&
\begin{pmatrix}
0 & \frac{-1}{\Pi_{12}}\\
\frac{-1}{\Pi_{12}}  & 0
\end{pmatrix},\\
\Pi_{11}\approx &
\int_{q_{1}}\frac{m}{q^{2}_{1}+q^{2}}=4 m \pi (-a_{IR} + a_{UV} + q {\rm atan}[a_{IR}/q] - q {\rm atan}[a_{UV}/q]),\\
\Pi_{12}\approx &
\int_{q_{1}}\frac{\Delta m^{2}}{q^{2}_{1}+q^{2}}=4 \Delta m^{2} \pi (-a_{IR} + a_{UV} + q {\rm atan}[a_{IR}/q] - q {\rm atan}[a_{UV}/q]).
\end{aligned}
\end{equation}
Then the correlation function reduced to
\begin{small}
\begin{equation} 
\begin{aligned}
&\chi(q,\Omega)=\frac{1}{\beta}
T_{1}(0)T_{2}(0)
G_{i}(-i\Omega,-q)G_{m}(i\Omega,q)G_{i}(0,0)G_{m}(0,0)
G_{i}(i\Omega,q)G_{i}(2i\Omega,2q)\\
=&
-\frac{1}{\beta(64 (q_{1IR} - q_{1UV})^{2} \Delta^{4} m^{4} \pi^{4} (i q^{2} + 2 m \Omega)^{2} (q^{4} + 
    i m q^{2} \Omega + 2 m^{2} \Omega^{2}) \eta^{2} (q_{1IR} - q_{1UV} - q {\rm atan}[\frac{q_{1IR}}{q}] + 
    q {\rm atan}[\frac{q_{1UV}}{q}])^{2})},
\end{aligned}
\end{equation}
\end{small}
and this correlation function satisfy the $f$-sum rule
\begin{small}
\begin{equation} 
\begin{aligned}
&\beta\int^{\infty}_{-\infty}\Omega{\rm Im}\chi(q,\Omega)d\Omega\\
=&
	\frac{1}{192 (q_{1IR} - q_{1UV})^{2} \Delta^{4} m^{6} \pi^{3} q^{4} \eta^{2} (2 q_{1IR} - 2 q_{1UV} + 
   q {\rm Arg}[1 - \frac{i q_{1IR}}{q}] - q {\rm Arg}[1 + \frac{i q_{1IR}}{q}] - 
   q {\rm Arg}[1 - \frac{i q_{1UV}}{q}] + q {\rm Arg}[1 + \frac{i q_{1UV}}{q}])^{2}}.
\end{aligned}
\end{equation}
\end{small}
Here to deal with the diffusion-like pole in $G_{i}(0,0),G_{m}(0,0)$,
we simply use the analytic continuation of Matsubara frequency and obtain $G_{i}(0,0)=G_{m}(0,0)\approx 1/(i\eta)$.
The integral range within this sum rule corresponds to the infinite bandwidth.
From Fig.,
the singular contribution to correlation function give rised by the small $\Omega$ at $q=0$ can be seen,
which means that, to explore the nonlocality of correlation function and the contribution of external momenta to the fluctuating polarons,
it is important to keep $\Omega$ finite, which is unlike the treatment we used above.
The singularity here also reveals the missing of correlation between two polaron,
as well as the instability when close to the critical point.
Note that the AL-type diagram discuss here is different from the standard AL diagram (as shown in Fig.9),
where, at temperature close to the critical one, the dependence on mode of Cooper channel
($Q,\mathcal{W}$) is only contained within the fluctuation propagator (i.e., the $T$-matrix),
but neglected during the calculations of the vertices\cite{Galitski V M}.
We note that due to the charge conjugation property,
the correlation function satisfies 
${\rm Re}\chi(q,\Omega)={\rm Re}\chi(q,-\Omega)={\rm Re}\chi(-q,\Omega)={\rm Re}\chi(-q,-\Omega)$
and ${\rm Im}\chi(q,\Omega)={\rm Im}\chi(-q,\Omega)=-{\rm Im}\chi(q,-\Omega)=-{\rm Im}\chi(-q,-\Omega)$
(since ${\rm sgn}[{\rm Im}\chi]=\mp$ corresponds to analytic continuation of $i\Omega=\Omega\pm i\eta$).

Now we know that the above correlation function satisfy the $f$-sum rule, but
what if the two impurities are correlated by the bubble with the external momenta origin from an external field instead of
the inner particle-hole bubble?
In this case, the correlation function (RPA-like bubble) becomes
\begin{equation} 
\begin{aligned}
\chi(q,\Omega)
=&\frac{1}{\beta}U^{2}
G_{i}(0,0)G_{i}(q,\Omega)\\
=&\frac{q^{2} \beta^{-1}}{2 m (\frac{q^{4}}{4 m^{2}}+\Omega^{2}) \eta},
\end{aligned}
\end{equation}
where $U$ denotes the relevant coupling constant.
We find this correlation function does not satisfies the $f$-sum rule,
i.e., we can nomore obtain a convergent result for the integral of $\int^{\infty}_{-\infty}\Omega{\rm Im}\chi(q,\Omega)d\Omega$
unless an infinite number of interaction rungs are taken into account.
This reveals that the bubble formed by these two impurities no longer behavior like the Schrodinger electrons.

\section{Free energy: mean-field part and fluctuation part}

Since the diagonal elements of $T$-matrices in Eq.(44) are zero in strong coupling limit,
the fluctuation contribution to the free energy is absent,
however, the mean-field contribution to the free energy is finite,
by treating the bipolaron system as a gapless collective bosonic propagator
(with $\chi(q,\Omega)$ the boson self-energy),
\begin{equation} 
\begin{aligned}
D_{bp}=\frac{1}{i\Omega_{bp}-(\varepsilon_{2q}+\chi(q,\Omega))},
\end{aligned}
\end{equation}
where $\varepsilon_{q}=\varepsilon_{Q}-\mu+\varepsilon_{Q'}-\mu=q^{2}/4m-2\mu$ (for zero mode $(Q,\mathcal{W})$),
and
this can be reduced to the summation of two bound-state energies $(-2E_{b})$ in strong-coupling case
when the chemical potential locates in the bottom of bound state band.
Similarly the polaron $T$-matrix can be reduced to the gapless free bosonic propagator at zero-temperature and tightly-bounded limit.

Using the the partition function for bosons
\begin{equation} 
\begin{aligned}
\mathcal{Z}=&\int \mathcal{D}(\phi^{*},\phi) e^{-S}\\
  =&\int \mathcal{D}(\phi^{*},\phi) e^{-\int^{\beta}_{0}d\tau\phi^{*}(\tau)D_{bp}^{-1}\phi(\tau)}\\
  =&\int \mathcal{D}(\phi^{*},\phi) e^{-\beta\phi^{*}D_{bp}^{-1}\phi}\\
  =&-{\rm det}\beta D^{-1}_{bp}\\
  =&-\beta^{2}((i\Omega)^{2}-(\varepsilon_{q}+\chi(q,\Omega))^{2})\\
  =&\beta^{2}(\Omega^{2}+(\varepsilon_{q}+\chi(q,\Omega))^{2}),
\end{aligned}
\end{equation}
where the relations $\varepsilon_{q}=\varepsilon_{-q}$, $\chi(q,\Omega)=\chi(-q,-\Omega)$ are used,
and we can then write
\begin{equation} 
\begin{aligned}
F_{{\rm MF}}
=&\frac{-1}{\beta V}\int_{q}\sum_{\Omega_{bp}}{\rm ln}\mathcal{Z}\\
=&\frac{-1}{\beta}\int_{q}[\beta(\varepsilon_{q}+\chi(q,\Omega))+ 2{\rm ln}[1+e^{-\beta(\varepsilon_{q}+\chi(q,\Omega))}]]\\
=&\frac{-2}{\beta}\int_{q}{\rm ln}[1+e^{-\beta(\varepsilon_{q}+\chi(q,\Omega))}]\\
\approx 
&-\frac{2}{\beta}4\pi 
{\rm ln}[1+ e^{2\mu\beta+(512 (q_{1IR} - q_{1UV})^{4} \Delta^{4} m^{8} \pi^{4} \Omega^{4} \eta^{2})^{-1}}] \frac{q^{3}}{3}
\bigg|_{q},
\end{aligned}
\end{equation}
where we have substracted the vacuum energy and the temperature-independent part of  thermodynamic potential
in third line of above equation,
and
we have used the identity
\begin{equation} 
\begin{aligned}
\int^{\beta^{2}\Omega_{bp}^{2}}_{1}\frac{dx}{x+\beta^{2}\varepsilon^{2}}+{\rm ln}[1+\beta^{2}\varepsilon^{2}]
={\rm ln}[\beta^{2}\varepsilon^{2}+\beta^{2}\Omega_{bp}^{2}].
\end{aligned}
\end{equation}
Note that if the collective propagator $D_{bp}$ in above calculation is replaced by complex scalar boson field propagator,
or with a boson kinetic term,
the mean field free energy should be obtained as\cite{Kapusta J I,Ohashi Y}
\begin{equation} 
\begin{aligned}
F_{{\rm MF}}
=&\frac{1}{\beta}\int_{Q}{\rm ln}[1-e^{-\beta(\varepsilon_{0}+\chi(q,\Omega))}].
\end{aligned}
\end{equation}

While in weak-coupling regime (with small momentum transfer $q$ and negligible Hartree-Fock terms) where the Thouless criterion is valid,
since only the diagonal elements of the matrix of pair propagator survive in the limit of $\Lambda\rightarrow\infty$,
the $T$-matrix becomes
\begin{equation} 
\begin{aligned}
{\bf T}(p)=&\frac{1}{(g_{b}^{-1}-\Pi_{22}(p))(g_{b}^{-1}-\Pi_{11}(p))}
\begin{pmatrix}
g_{b}^{-1}-\Pi_{11}(p) & \phi(0)\\
\phi(0)  & g_{b}^{-1}-\Pi_{22}(p)
\end{pmatrix}\\
=&
\begin{pmatrix}
\frac{1}{(g_{b}^{-1}-\Pi_{22}(p))} &\frac{ \phi(0)}{(g_{b}^{-1}-\Pi_{22}(p))(g_{b}^{-1}-\Pi_{11}(p))}  \\
\frac{ \phi(0)}{(g_{b}^{-1}-\Pi_{22}(p))(g_{b}^{-1}-\Pi_{11}(p))}  & \frac{1}{(g_{b}^{-1}-\Pi_{11}(p))}
\end{pmatrix}\\
\approx &
\begin{pmatrix}
g_{b} &\phi(0)g_{b}^{2}  \\
\phi(0)g_{b}^{2}    & g_{b}
\end{pmatrix}\\.
\end{aligned}
\end{equation}
For mode $(Q,\mathcal{W})=(0,0)$ in the center-of-mass framework, where $\Pi_{11}=\Pi_{22}$,
the free energy can be obtained as
\begin{equation} 
\begin{aligned}
F_{{\rm Flu}}
=&-\frac{1}{\beta}\sum_{\Omega}\int_{q}{\rm Tr}{\rm ln}[1-g_{b}\mathcal{C}(q,\Omega)]\\
=&-\frac{1}{\beta}\sum_{\Omega}\int_{q}{\rm Tr}{\rm ln}[g_{b}(g_{b}^{-1}-\mathcal{C}(q,\Omega))]\\
=&-\frac{1}{\beta}\sum_{\Omega}\int_{q}{\rm Tr}{\rm ln}[g_{b}\Sigma_{AL}^{-1}]\\
=&-\frac{1}{\beta}\sum_{\Omega}\int_{q}{\rm Tr}{\rm ln}[g_{b}\chi^{-1}(q,\Omega)]\\
=&-\frac{1}{\beta}\sum_{\Omega}\int_{q}{\rm ln}[g_{b}\chi_{11}^{-1}(q,\Omega)]
-\frac{1}{\beta}\sum_{\Omega}\int_{q}{\rm ln}[g_{b}\chi_{22}^{-1}(q,\Omega)],
\end{aligned}
\end{equation}
where $\mathcal{C}(q,\Omega)$ is the correlation function whose expression should not contains any coupling term,
and  $\Sigma_{AL}$ is the AL-type self-energy, which equals to $\chi(q,\Omega)$ in Eq.(42) since we can treat the bipolaron as a
composite boson here.
Then for the first ($\Pi$) and second ($\Pi'$) polarons, the pair propagators can be obtained (in the limit of small $q_{1}$) as
\begin{equation} 
\begin{aligned}
\Pi_{11}=&\frac{1}{\beta}\sum_{n'}\int_{q_{1}}
G_{11}(-q_{1})G_{11}(q_{1})\\
=&
\frac{{\rm csc}[\frac{\sqrt{-\Delta^{2} m^{2}}}{
  2 m \beta^{-1}}]^{2} (\Delta^{2} m -
   \sqrt{-\Delta^{2} m^{2}} \beta^{-1} 
   {\rm sin}[\frac{\sqrt{-\Delta^{2} m^{2}}}{m \beta^{-1}}])}{8 (\Delta^{2} m \beta^{-1})}
+O(q_{1}^{3}),\\
\Pi'_{11}=&\frac{1}{\beta}\sum_{n'}\int_{q_{1}}
G_{11}(q-q_{1})G_{11}(q+q_{1})\\
=&
\frac{1}{4} m (-(
\frac{
    2 (\Delta^{2} m^{2} \beta^{-1} - i m \Omega q^{2} \beta^{-1} - 
       \Omega a) {\rm cot}[\frac{
      2 m^{2} \Omega \beta^{-1} + a}{
      4 m^{2} \beta^{-2}}]}{
    m \Omega (4 \Delta^{2} m^{2} \beta^{-1} + q^{4} \beta^{-1} - 
       2 \Omega a)}) \\
&- \frac{
   2 (\Delta^{2} m^{2} \beta^{-1} - i m \Omega q^{2} \beta^{-1} + 
      \Omega a) {\rm cot}[
     \frac{\Omega}{2 \beta^{-1}} - \frac{a}{
      4 m^{2} \beta^{-2}}]}{
   m \Omega (4 \Delta^{2} m^{2} \beta^{-1} + q^{4} \beta^{-1} + 
      2 \Omega a)})
+O(q_{1}^{3}),\\
\Pi_{12}=&\frac{1}{\beta}\sum_{n'}\int_{q_{1}}
G_{12}(-q_{1})G_{12}(q_{1})\\
=&
\frac{{\rm csc}[\frac{\sqrt{-\Delta^{2} m^{2}}}{
  2 m \beta^{-1}}]^{2} 
  (-\Delta^{2} m +    \sqrt{-\Delta^{2} m^{2}} \beta^{-1} {\rm sin}[\frac{\sqrt{-\Delta^{2} m^{2}}}{m \beta^{-1}}])}
 {8 \Delta^{2} m \beta^{-1}}
+O(q_{1}^{3}),\\
\Pi'_{12}=&\frac{1}{\beta}\sum_{n'}\int_{q_{1}}
G_{12}(q-q_{1})G_{12}(q+q_{1})\\
=&
	  (-(\frac{\Delta^{2} m^{2} \beta^{-1} {\rm cot}[\frac{
     2 m^{2} \Omega \beta^{-1} + a}{
     4 m^{2} \beta^{-2}}]}{
   2 \Omega (-4 \Delta^{2} m^{2} \beta^{-1} - q^{4} \beta^{-1} + 
      2 \Omega a)}) + \frac{
  \Delta^{2} m^{2} \beta^{-1} {\rm cot}[
    \frac{\Omega}{2 \beta^{-1}} - \frac{a}{4 m^{2} \beta^{-2}}]}{
  2 \Omega (4 \Delta^{2} m^{2} \beta^{-1} + q^{4} \beta^{-1} + 
     2 \Omega a)})
+O(q_{1}^{3}),
\end{aligned}
\end{equation}
where $a=\sqrt{-m^{2} (4 \Delta^{2} m^{2}+ q^{4}) \beta^{-2}}$,
$\Omega_{1}=2n'\pi\beta^{-1}$.
We have
\begin{equation} 
\begin{aligned}
\chi_{11}(q,\Omega)=&
\frac{1}{\beta}
\Pi_{11}\Pi'_{11}G_{11}(-q_{1})G_{11}(q_{1})G_{11}(0)G_{11}(0)G_{11}(p-q_{1})G_{11}(p+q_{1}),\\
\chi_{22}(q,\Omega)=&
\frac{1}{\beta}
\Pi_{11}\Pi'_{11}G_{11}(q_{1})G_{11}(-q_{1})G_{11}(0)G_{11}(0)G_{11}(-p+q_{1})G_{11}(-p-q_{1}).
\end{aligned}
\end{equation}
Then by substituting the above equation into Eq.(69),
we can obtain the fluctuation part of free energy.

In this regime,
the fluctuation part of free energy (thermodynamic potential)
can also be obtained through the dynamical pair susceptibility of the bipolaron,
base on the effective interaction $g_{b}^{2}D_{0}(\delta q,\delta\Omega)$ between two impurities
(instead of two $T$-matrices; see Fig.11)
\begin{equation} 
\begin{aligned}
F_{{\rm Flu}}=-\frac{1}{\beta}\sum_{\Omega}\int_{q}{\rm Tr}{\rm ln}[g_{b}^{2}D(\delta q,\delta\Omega)
\Pi_{m}(q,\Omega)\Sigma^{-1}_{AL}(q,\Omega)],
\end{aligned}
\end{equation}
with the dressed (and massive) boson propagator of particle-hole excitation reads
\begin{equation} 
\begin{aligned}
D=\frac{1}{\beta}\sum_{\delta \Omega}\int_{\delta q}
D(\delta \Omega,\delta q),
\end{aligned}
\end{equation}
whose matrix form is
\begin{equation} 
\begin{aligned}
D^{-1}(\delta \Omega,\delta q)=&
\begin{pmatrix}
i\delta\Omega-\varepsilon_{\delta q} & 0\\
0 & i\delta\Omega+\varepsilon_{-\delta q}
\end{pmatrix}
-
\begin{pmatrix}
-g_{b}^{2}\uppi_{11}(\delta \Omega,\delta q) & m_{\phi}\\
m_{\phi}^{*} & -g_{b}^{2}\uppi_{22}(\delta \Omega,\delta q) 
\end{pmatrix}\\
=&
\begin{pmatrix}
i\delta\Omega-\varepsilon_{\delta q}-g_{b}^{2}\uppi_{11}(\delta \Omega,\delta q) & g_{b}\phi(\delta q)\\
g_{b}\phi^{*}(\delta q) & i\delta\Omega+\varepsilon_{-\delta q}-g_{b}^{2}\uppi_{22}(\delta \Omega,\delta q) 
\end{pmatrix},
\end{aligned}
\end{equation}
Thus 
\begin{equation} 
\begin{aligned}
D_{11}(\delta \Omega,\delta q)=&
\frac{i\delta\Omega+\varepsilon_{-\delta q}-g_{b}^{2}\uppi_{22}(\delta \Omega,\delta q) }
{(i\delta\Omega-\varepsilon_{\delta q}-g_{b}^{2}\uppi_{11}(\delta \Omega,\delta q) )
(i\delta\Omega+\varepsilon_{-\delta q}-g_{b}^{2}\uppi_{22}(\delta \Omega,\delta q) )-g_{b}^{2}|\phi(\delta q)|^{2}}\\
D_{12}(\delta \Omega,\delta q)=&
\frac{g_{b}^{2}|\phi(\delta q)|^{2}}
{(i\delta\Omega-\varepsilon_{\delta q}-g_{b}^{2}\uppi_{11}(\delta \Omega,\delta q) )
(i\delta\Omega+\varepsilon_{-\delta q}-g_{b}^{2}\uppi_{22}(\delta \Omega,\delta q) )-g_{b}^{2}|\phi(\delta q)|^{2}}.
\end{aligned}
\end{equation}
where we use the nonrelativistic boson dispersion  $ \varepsilon_{\delta q}=\frac{\delta q^{2}}{2m_{\phi}}+m_{\phi}$.
Note that the summation over Matsubara frequency $\delta\Omega$ of $D_{11}(\delta \Omega,\delta q)$
is divergent, and thus requires a convergence factor $e^{i\delta\Omega \eta}(\eta\rightarrow 0)$ 
or the momentum cutoff.

The AL-diagram-type self-energy $\Sigma_{AL}(q,\Omega)$ 
\begin{equation} 
\begin{aligned}
\Sigma_{AL}(q,\Omega)
=&(g_{b}^{2}D_{0}\Pi_{m}+(g_{b}^{2}D_{0}\Pi_{m})^{2}\Pi_{i}(q,\Omega)+(g_{b}^{2}D_{0}\Pi_{m})^{3}\Pi_{i}^{2}(q,\Omega)+\cdot\cdot\cdot)
G(\frac{Q}{2}-q)G(\frac{Q'}{2})    G(\frac{Q}{2})G(\frac{Q'}{2}+q)\\
=&
\frac{G(\frac{Q}{2}-q)G(\frac{Q'}{2})    G(\frac{Q}{2})G(\frac{Q'}{2}+q)}
{(g_{b}^{2}D_{0}\Pi_{m})^{-1}-\Pi_{i}(q,\Omega)} ,
\end{aligned}
\end{equation}
with the two-particle correlations defined as
\begin{equation} 
\begin{aligned}
\Pi_{11m}(q,\Omega)
=&\frac{1}{\beta}\sum_{\delta\Omega}\int_{\delta q}
G_{11}(\frac{Q}{2}+\delta q)G_{11}(\frac{Q'}{2}-\delta q)\\
=&\frac{1}{\beta}\sum_{\delta\Omega}\int_{\delta q}
G_{11}(\delta q)G_{11}(q-\delta q),\\
\Pi_{11i}(q,\Omega)
=&\frac{1}{\beta}\sum_{\delta \Omega}\int_{\delta q}
G_{11}(\frac{Q}{2}-\delta q)G_{11}(\frac{Q'}{2}+\delta q)\\
=&\frac{1}{\beta}\sum_{\delta \Omega}\int_{\delta q}
G_{11}(-\delta q)G_{11}(q+\delta q),
\end{aligned}
\end{equation}
where $\delta q$ is the momentum transfer of each interaction rung,
and we can define a renormalized $T$-matrix within the above self-energy, 
\begin{equation} 
\begin{aligned}
T_{{\rm ren}}=
\frac{1}
{(g_{b}^{2}D_{0}\Pi_{m})^{-1}-\Pi_{i}(q,\Omega)},
\end{aligned}
\end{equation}
as shown in Fig.11.
We also shown the real part of above correlation functions $\Pi_{m}$ and $\Pi_{i}$ in Fig.12.
Although not shown, it is also found that ${\rm Im}\Pi_{i}(q)\approx {\rm Im}\Pi_{m}(q)\approx \delta(q)$ at $\Omega=0$
(while for analytically continued one we have ${\rm Im}\Pi_{i}(q)\approx -{\rm Im}\Pi_{m}(q)\approx \delta(q)$ at $\Omega=0$),
which implies that the spectral functions of pair propagator $A(\Omega)=-{\rm Im}\Pi(\Omega+i\eta)$ are approximately the delta function at $(q,\Omega)=(0,0)$,
i.e., in the absence of momentum transition (or bipolaron),
and at low-temperature limit,
the correlator operator in imaginary time-domain reads
$D(\tau)=\int^{\infty}_{-\infty}d\Omega\frac{e^{-\tau\Omega}}{1-e^{-\beta\Omega}}A(\Omega)
\approx \int^{\infty}_{-\infty}d\Omega e^{-\tau\Omega}A(\Omega)$.
While for large value of $(q,\Omega)$,
the spectral function will exhibits strong non-Fermi-liquid feature.
Then by substituting the above equations into Eq.(72),
we can obtain the fluctuation part of free energy.

Next we consider the low-temperature limit where the quantum fluctuation provides an important contribution to the 
fluctuation part of free energy.
In this limit, for zero $\delta\Omega$, 
we have
\begin{equation} 
\begin{aligned}
D_{11}
=&\int_{\delta q}D_{11}(\delta q) 
=&-i \sqrt{-\frac{1}{2} + \frac{1}{\sqrt{2}}} \pi=-D_{22},\\
D_{12}
=&\int_{\delta q}D_{12}(\delta q) 
=&i \sqrt{\frac{1}{2} + \frac{1}{\sqrt{2}}} m_{\phi} \pi=D_{21},\\
\end{aligned}
\end{equation}
where the nonrelativistic boson dispersion defined above is used.
Thus to keep the coupling $(g_{b}^{2}D_{0}\Pi_{m})$ to be the attractive-type one (i.e., be negative),
the pair propagator $\Pi_{m}$ must be positive (see Fig.).
The mean field term $\phi(0)$ is not being contained in the boson mass term $m_{\phi}$,
and in zero-temperature limit,  the boson mass is much smaller than the fermion mass, $m_{\phi}\ll m$.
Since in weak-coupling limit, $q\rightarrow 0$ and $g_{b}\rightarrow 0$ (thus $\Delta\rightarrow 0$),
through complex calculations,
we obtain the contribution of quantum fluctuation to the free energy as
\begin{equation} 
\begin{aligned}
F_{{\rm Q-Flu}}
\approx &\int_{q}{\rm ln}[\frac{2i\eta \Delta^{6}m^{3}}{q^{6}}(1+g_{b}^{2}\mathcal{F}_{1}(\Omega,\Delta))]
                        +\int_{q}{\rm ln}[\frac{2i\eta m_{\phi}^{6}m^{3}}{q^{6}}(1-g_{b}^{2}\mathcal{F}_{2}(\Omega,\Delta))]\\
\approx &\int_{q}{\rm ln}[e^{iqr}\frac{2i\eta \Delta^{6}m^{3}}{q^{6}}(1+g_{b}^{2}\mathcal{F}_{1}(\Omega,\Delta))]
                        +\int_{q}{\rm ln}[e^{iqr}\frac{2i\eta m_{\phi}^{6}m^{3}}{q^{6}}(1-g_{b}^{2}\mathcal{F}_{2}(\Omega,\Delta))]\\
=&\frac{i (6 (\gamma + {\rm ln}[-i r]) + 
    {\rm ln}[2 i \Delta^{6} (1 + \mathcal{F}_{1} g_{b}^{2}) m^{3} \eta])}{r} + 
 \frac{ i (6 (\gamma + {\rm ln}[-i r]) + 
    {\rm ln}[-2 i \Delta^{6} (-1 + \mathcal{F}_{2} g_{b}^{2}) m^{3} \eta])}{r}.
\end{aligned}
\end{equation}
where $\mathcal{F}_{1}$ and $\mathcal{F}_{2}$ are the functions which are independent of $q$,
and $\gamma\approx 0.577216$ is the Euler's constant.
A convergent factor $e^{iqr}(r\ll 1)$ is used in second line, instead of a finite UV cutoff,
since the UV cutoff $\Lambda\rightarrow \infty$ in weak coupling limit.
Here $r$ is an imaginary quantity with ${\rm Im}r>0$,
thus the above quantum fluctuation part of free-energy is real,
in contrast to that in classical limit ($\beta^{-1}\gg\Omega$).
Note that the quantum fluctuation is suppressed in high space-dimension ($d>2$) and also by the finite excitation (bosonic) mass.

Since the critical temperature ($T_{c}\sim\Lambda^{-1}$) is very low in weak-coupling limit,
the Thouless condition can be satisfied for the above renormalized $T$-matrix (at zero-temperature) as
\begin{equation} 
\begin{aligned}
(g_{b}^{2}D_{0}\Pi_{m})^{-1}-\Pi_{i}(q,\Omega)=0.
\end{aligned}
\end{equation}
Thus we can see that in weak-coupling limit,
as the critical coupling $(g_{b}^{2}D_{0}\Pi_{m})\rightarrow 0$,
the bound state as well as the resonance pole corresponds to the divergent $|\Pi_{i}(q,\Omega)|$,
which corresponds to the critical $\Omega_{c}$ as shown in Fig.10.
The critical bosonic frequency $\Omega_{c}$ is obtained by solving $|\Pi_{i}(q,\Omega)^{-1}|=0$,
and it turns out that $\Omega_{c}$ is infinitely closes but not equal to zero.
We note that, while in RG scheme, the
 four-fermion coupling will nomore be a constant but diverge exponentially at critical temperature $T_{c}$ (with scale parameter $\ell\rightarrow\infty$)\cite{Throckmorton R E},
i.e.,
$g_{GR}^{-1}\sim e^{\ell}/T_{c}$, $dg_{RG}/d\ell \sim e^{\ell}T_{c}
\sim -g'_{RG}$ (in the case of $z=1$).
Such exponent divergent of  coupling can also be found within RPA as the Landau pole\cite{Sushkov A B,Jian S K} for
the relativistic fermions, like the massless Dirac fermions.

As the above solved renormalized $T$ matrix can be rewritten as
\begin{equation} 
\begin{aligned}
&{\bf T}_{{\rm ren}}=\frac{1}{{\rm Det}}\\
&\begin{pmatrix}
(g_{b}^{2}D_{0}\Pi_{11m}^{--})^{-1}-\Pi_{11}^{--} & \Pi_{12m}^{++} & 0 & \Pi_{12i}^{++}\\
\Pi_{21m}^{++} & (g_{b}^{2}D_{0}\Pi_{11m}^{++})^{-1}-\Pi_{11}^{--}  & \Pi_{12i} & 0\\
0 & \Pi_{21i} & (g_{b}^{2}D_{0}\Pi_{11m}^{--})^{-1}-\Pi_{11}^{++} & \Pi_{12m}^{++}\\
\Pi_{21i} & 0 & \Pi_{12m}^{++} & (g_{b}^{2}D_{0}\Pi_{11m}^{++})^{-1}-\Pi_{11}^{++}
\end{pmatrix},
\end{aligned}
\end{equation}
where the superscript $(\pm,\pm)$ corresponds to $(\pm q,\pm\Omega)$.
${\rm Det}$ is the determinant term of the above matrix.
Thus the delta-type imaginary part (undamped) of pair propagator $\Pi_{i}$ as discussed in above at $(q,\Omega)=(0,0)$
corresponds to ${\rm Det}=0$.
While for nonvanishing mode $(q,\Omega)$,
the relation ${\rm Det}=0$ can be reduced to, at critical temperature,
$(\Pi_{12m}^{2}(q,\Omega)-\Pi_{12i}^{2}(q,\Omega))^{2}=0$,
i.e., 
$\Pi_{12m}(q,\Omega)=\Pi_{12i}(q,\Omega)$ or
$\Pi_{12m}(q,\Omega)=-\Pi_{12i}(q,\Omega)$.

Next we consider the case that
the impurities are coupled to the bath of boson field,
in which case the single boson propagator with the denominator of  renormalized $T$-matrix (Eq.(73))
should be replaced by the boson field propagator (Eq.(11)).
Unlike the fermionic bath coupling which connects the fermion and boson propagators,
the bosonic bath coupling connects the fermion propagators (particle-hole excitations) diagrammatically\cite{Joshi D G},
\begin{equation} 
\begin{aligned}
D^{-1}_{0}(\delta \Omega,\delta q)=&
\begin{pmatrix}
-i\delta\Omega+\delta\Omega^{2}+\delta q^{2}+m_{\phi}^{2} & 0\\
0 & -(i\delta\Omega+\delta\Omega^{2}+\delta q^{2}+m_{\phi}^{2})
\end{pmatrix},
\end{aligned}
\end{equation}
thus we have (still at zero $\delta\Omega$), in both particle-hole symmetry and asymmetry cases (in absence of complex linear frequency term),
\begin{equation} 
\begin{aligned}
D_{0}=\int_{\delta q}D_{0}(\delta \Omega,\delta q)=\frac{\pi}{\sqrt{m_{\phi}}}.
\end{aligned}
\end{equation}
The resulting free energy $F_{{\rm Q-flu}}$ has the same form with Eq.(74),
but with a different $F$ function which depends on the boson mass term $m_{\phi}$.
It is important to note that the above equation fails for the particle-particle (anomalous) order parameters in which case the $m_{\phi}$ is purely real.

The bosonic fluctuation part of free energy at finite temperature can also be obtained by the Nozieres-Schmitt-Rink (NSR) approximation,
\begin{equation} 
\begin{aligned}
F_{NSR}=&
\int_{q}\int^{\infty}_{-\infty}
\frac{d\Omega}{2\pi i}\frac
{{\rm ln}[g_{b}^{2}D_{0}\Pi_{m}(\Sigma_{AL}^{R}(q,\Omega))^{-1}]-{\rm ln}[g_{b}^{2}D_{0}\Pi_{m}(\Sigma_{AL}^{A}(q,\Omega))^{-1}]}
{e^{\beta\Omega}-1}\\
=&\int_{q}\int^{\infty}_{-\infty}
\frac{d\Omega}{2\pi}
\frac{-2\delta(q,\Omega)}
{e^{\beta\Omega}-1},
\end{aligned}
\end{equation}
where $\Sigma_{AL}^{R}(q,\Omega)$ and $\Sigma_{AL}^{A}(q,\Omega)$ are the retarded and advanced AL-type self-energy, respectively.
The phase shift $\delta(q,\Omega)$, which $\propto g_{b}$,
is defined through the relation
\begin{equation} 
\begin{aligned}
\frac{g_{b}^{2}D_{0}\Pi_{m}(\Sigma_{AL}^{R/A}(q,\Omega))^{-1}}{g_{b}^{2}D_{0}\Pi_{m}(\Sigma_{AL}(q,\Omega))^{-1}}=e^{\mp i\delta(q,\Omega)}.
\end{aligned}
\end{equation}

\section{Summary}

Unlike the fermionic mass, the order-parameter mass term will not directly gives rise to nonadiabatic (or nonrelativistic) dynamics,
and the non-Fermi-liquid behavior in bipolaron system will further requires strong quantum fluctuations and
the large number of gapless bosonic excitations in the low-energy limit (and with an UV cutoff in frequency $\Lambda\rightarrow\infty$).
The constant coupling $g_{b}$ in this paper is assumed indepedent of the momentum,
but it is a nonlocal term (suppressed by $1/N$) and can be turned to zero logarithmically in the $\Lambda\rightarrow\infty$ limit,
as it is an irrelevant parameter under renormalization,
and the without effect on the low-energy physics in weak-coupling limit.
In low-energy limit with weak-coupling,
the linear frequency term in fermion field propagator can be ignored\cite{Lee S S},
then the bipolaron can still exists as a propagating mode even in electron-hole symmetry case
due to the existence of frequency quadratic term within the boson field propagator.

The properties of bipolaron are mainly studied by using the Green's function method,
and from the self-energies (the normal one), it can be seen that the particle-hole symmetry is absent
($-\Sigma^{*}(-\mathcal{W})=\Sigma(\mathcal{W})$).
We also found that the pole structure (resonance) can be found in polaron and bipolaron self-energies
even above the critical temperature (where the anomalous self-energy vanishes), but is absent
in the bosonic self-energies.
The mean-field contribution and fluctuation contribution to the free energy are studied in this paper,
where the finite-temperature fluctuation part can also be obtained by the NSR approximation,
while the quantum fluctuation part is dominate only in the low-temperature limit and low-energy limit (with weak-coupling)
with large number of massless boson modes ($m_{\phi}=0$).
This also reveals that strong quantum fluctuation can also be caused by the virtual particle-hole excitations\cite{Lee S S},
on a well-defined fermi surface (which only exists in fermi-liquid state),
and this effect is absent in both the relativistic quantum field theory (QFT) and nonrelativistic NSR theory.

Although the role of large-$N$ expansion does not shown in this paper, it is indeed usefull in high-temperature
strong-coupling regime (in contrast to the tightly bounded regime)
where $E_{b}^{-1}\gg 1$,  by using the perturbative expansion in terms of $1/N$ in large-$N$ limit.
The fermi-liquid behavior of the particle-particle correlation function is also guaranteed in the large-$N$ limit,
until the long-range interaction (with gapless boson field) or the strong quantum fluctuation at low-energy limit appear.
As long as all the internal and external momenta are close enough to the fermi surface,
the order of Feynman diagrams to leading order in $1/N$ can be obtained.
We replot in Fig. the bipolaron diagrams in Fig.1 and the AL-type one in Fig.9(a)
Accoding to the procedure of Ref.\cite{Lee S S} and the facts that each external fermion line contributes a order of $N^{1/2}$,
(thus each external bosonic line contributes a order of $N$),
and each vertex contributes $\frac{1}{\sqrt{N}}$,
we can obtain that
the bipolaron diagram with external fermion lines is of the order of $N^{-1}N^{-2}N^{2}=1/N$ (Fig.13(d-e)),
and the AL-type one with external boson lines is also of the order of $N^{-1}N^{2}N^{-2}=1/N$ (Fig.13(h-i)),
as summarized in Fig.13.
This conclusion is valid to any order of bare coupling $g_{b}$,
i.e., no matter how large the number of loops within the diagram is,
and it is found that this conclusion has some difference to the diagrams which is constituted by only the particle-hole pairs\cite{Lee S S}
(i.e., the multi-loop fermion RPA diagram).

\section{Appendix.A: Green's function}

Firstly the Matsubara Green's function is defined as
\begin{equation} 
\begin{aligned}
G(i\omega_{n})=G(i(2n+1)\pi/\beta)=\int^{\beta}_{0}
d\tau G(\tau)e^{i\omega_{n}\tau},
\end{aligned}
\end{equation}
or in terms of the retarded Green's function as
\begin{equation} 
\begin{aligned}
G(i\omega_{n})
=&\int^{\infty}_{0}
dt G^{R}(t)e^{i\omega_{n}\tau}\\
=&\int^{\infty}_{0}
dt [-i\theta(t)\langle\mathcal{T}\{c(t),c^{\dag}(0)\}\rangle] e^{i\omega_{n}\tau}\\
=&\int^{\infty}_{0}
dt [-i\theta(t)\langle\mathcal{T}\{c(t),c^{\dag}(0)\}\rangle] e^{-\omega_{n}t},
\end{aligned}
\end{equation}
where
$\mathcal{T}$ is the time ordering operator.
For a interacting system with long-time evolution described by wavefunction $\psi$,
we have $\psi(t)=U^{t}_{0}\psi(0)$,
where we assume the interaction (or perturbations)
starts at an initial time $t=0$.
Here the evolution operator $U^{t}_{0}$ has a Dyson-type formula
\begin{equation} 
\begin{aligned}
U^{t}_{0}
=&1-i\int^{t}_{0}H(t')U^{t'}_{0}dt'\\
=&1-i\int^{t}_{0}H(t')dt'-\int^{t}_{0}dt'\int^{t'}_{0}dt''H(t')H(t'')+\cdot\cdot\cdot,
\end{aligned}
\end{equation}
or in terms of time ordering operator the above formula can be rewritten as
\begin{equation} 
\begin{aligned}
U^{t}_{0}
=&\mathcal{T}exp(-i\int^{t}_{0}dt'H(t'))\\
=&exp(-i\int^{t}_{0}dt'H(t')+\frac{1}{2}\int^{t}_{0}dt'\int^{t'}_{0}dt''[H(t''),H(t')]\\
 &-\frac{1}{4}\int^{t}_{0}dt'\int^{t'}_{0}dt''\int^{t''}_{0}dt'''[H(t'''),[H(t''),H(t')]]).
\end{aligned}
\end{equation}
For an noninteracting contour Keldysh-type Green's function,
\begin{equation} 
\begin{aligned}
G^{K}(t)=&-i\langle \mathcal{T}c(t)c^{\dag}(0)\rangle\\
        =&-i\langle c(t)c^{\dag}(0)\rangle\\
        =&i\langle \mathcal{T}c^{\dag}(0)c(t)\rangle.
\end{aligned}
\end{equation}
The greater Green's functions in frequency domain
corresponds to the above Keldysh-type Green's function reads
\begin{equation} 
\begin{aligned}
G(\omega_{n})
        =\frac{2i\eta(N(\omega)-1)}{(\omega-\varepsilon)^{2}+\eta^{2}},
\end{aligned}
\end{equation}
where $\eta$ is the relaxation rate,
and $N(\omega)$ is the frequency-dependent distribution function,
which is equivalent to the Fermi distribution function
for a delta-type spectral function,
e.g., the one in an Anderson (fixed) impurity model which exhibits a sharp Kondo resonance.
Note that even the slightly smeared (not strictly delta-type) spectral function
can be treated as a signature of Fermi-liquid state,
with a well-defined Fermi surface.
The imaginary part of fermion field and boson field propagators in imaginary time domain are shown in Fig.13.

\section{Appendix.B: Coupling constant}

$g_{b}$ is the bare polaronic coupling between a single impurity and the majority particle,
which is momentum-independent and can be obtained by the single polaron binding energy as
\begin{equation} 
\begin{aligned}
g_{b}^{-1}=&-\int\frac{d^{3}q}{(2\pi)^{3}}
\frac{1}{E_{b}+\varepsilon_{-q}+\varepsilon_{q}}\\
=&-4 m \pi (\Lambda_{q} - 
   \sqrt{E_{b}} \sqrt{m} {\rm atan}[\frac{\Lambda_{q}}{\sqrt{E_{b}} \sqrt{m}}]).
\end{aligned}
\end{equation}

Note that in the strong coupling limit
the perturbation calculation can still be used here through the large-flavor number expansion,
where the flavor number $N$ is brought by each particle-hole (or particle-particle) loop
and $1/N$ is brought by the vertices,
and the above response function (Eq.()) can then still be obtained by the result to
leading order in $1/N$.
Note that the lowest order in $1/N$ expansion is equivalent to the lowest order $T$-matrix calculation
or the Nozieres–Schmitt-Rink theory.
Since the dressed bosonic propagator $D(q,\Omega)$ describes the strengh of interaction 
between two polarons through the excitations in majority component,
and $D(q,\Omega)\sim 1/N \rightarrow 0$ as $N\rightarrow\infty$ (in the mean time the dynamical critical exponent 
$z\rightarrow 1$),
the interaction between two polarons remains weak no matter how large the $g_{b}$ is.
The propagator of scalar bosonic mode as
well as the beyond-mean-field fluctuation also vanishes at mean-field level with $N\rightarrow\infty$.

At finite temperature with weak coupling,
the above bare propagator $D_{0}(q,\Omega)$ can be regarded as adiabatic propagator
(in Migdal-Eliashberg approximation)
with $\Omega\ll\mu_{m}$ and $q\ll k_{F}$, while the dressed one $D(q,\Omega)$ considers the non-adiabatic 
contribution and thus contains decay effect with the particle-hole continuum.
The bosonic propagator here is also different from the Coulomb propagator which is dimension-dependent
($D_{0}\sim 1/q$ in 2D and $D_{0}\sim 1/q^{2}$ in 3D).
Note that, here we can certainly extend the above self-energy expression to make it 
contains infinite number of loops
theoretically,
which is similar to the ladder approximation appearing in the self-energy of single-polaron.
That leads to a larger inverse quasiparticle lifetime compared to the inverse disorder lifetime,
${\rm Im}\Sigma>\frac{1}{\tau_{diso}}=2\pi n_{i}U^{2}\rho(\omega)$ where $U$ denotes the disorder potentials. 

\section{Appendix.C: Relation to fluctuations in other systems}

It is obviously that the coupling of particle-hole bubble to a particular fluctuating order-parameter field
is essential to the formation of bipolaron here.
We note that this is similar to the case of spin-fluctuation-induced pairing gap in superconductors\cite{Sknepnek R},
but unlike the case in the ultra-cold Fermi gases\cite{Mukherjee B},
where the fluctuations 
(usually the particle-particle fluctuation)
will supresses the stable-pair formation
by, for example,
lowers the critical temperature and increases the pairing gap,
i.e., the system is being rendered to the BEC (dimer) regime with Yukawa coupling 
instead of the mean field BCS regime with the Ruderman-Kittel-Kasuya-Yosida coupling\cite{Suchet D}.

\section{Appendix.D: Derivation of Eq.(41)}

As diagrammatically shown in Fig.,
the self-energy of bipolaron can be written,
in terms of the convolution of propagators of impurities and excited particle and hole, as
\begin{equation} 
\begin{aligned}
\Sigma(p_{1},p_{2},\omega_{1},\omega_{2})
=&g_{b}\uppi(q,\Omega)g_{b}\\
&
+g_{b}\uppi(q,\Omega)g_{b}^{2}\int_{q_{1}}G(p_{2}+q_{1})G(k+q-q_{1})
+g_{b}^{2}\int_{q_{1}}G(p_{1}-q_{1})G(k+q_{1})\uppi(q,\Omega)g_{b}\\
&
+g_{b}^{2}\int_{q_{1}}G(p_{1}-q_{1})G(k+q_{1})\uppi(q,\Omega)g_{b}^{2}\int_{q_{1},q_{2}}G(p_{2}+q_{1})G(k+q-q_{1})\\
&
+g_{b}\uppi(q,\Omega)g_{b}^{3}\int_{q_{1},q_{2}}G(p_{2}+q_{2})G(k+q-q_{2})G(p_{2}+q_{1})G(k+q-q_{1})\\
&
+g_{b}^{3}\int_{q_{1},q_{2}}G(p_{1}-q_{2})G(k+q_{2})G(p_{1}-q_{1})G(k+q_{1})\uppi(q,\Omega)g_{b}\\
&
+O(g_{b}^{5})\\
=&(g_{b}+g_{b}^{2}\int_{q_{1}}G(p_{1}-q_{1})G(k+q_{1})\\
&+g_{b}^{3}\int_{q_{1},q_{2}}G(p_{1}-q_{2})G(k+q_{2})G(p_{1}-q_{1})G(k+q_{1})+O(g_{b}^{4}))\\
&\uppi(q,\Omega)
(g_{b}+g_{b}^{2}\int_{q_{1}}G(p_{2}+q_{1})G(k+q-q_{1})\\
&+g_{b}^{3}\int_{q_{1},q_{2}}G(p_{2}+q_{2})G(k+q-q_{2})G(p_{2}+q_{1})G(k+q-q_{1})+O(g_{b}^{4}))\\
=&\int_{q,\Omega}\frac{g_{b}}{1-g_{b}\int_{q_{1}}G(p_{1}-q_{1})G(k+q_{1})}\frac{g_{b}}{1-g_{b}\int_{q_{1}}G(p_{2}+q_{1})G(k+q-q_{1})}
\uppi(q,\Omega)\\
=&\int_{q,\Omega}T_{1}(p_{1},\omega_{1})\uppi(q,\Omega)T_{2}(p_{2},\omega_{2}).
\end{aligned}
\end{equation}
Note that all the Green's functions containing $k$ are the propagator of majority particle,
and all the Green's functions containing $p$ are the propagator of impurity.
Similarly, for the tri-polaron, we can obtain the self-energy
as $T_{1}T_{2}T_{3}\uppi_{3}$ where $\uppi_{3}$ is the density-density-density correlation,
and usually, such tri-polaron mode is related to the $\phi^{3}$ term as well as the six-point vertex.
\end{large}
\renewcommand\refname{References}

\clearpage
\begin{figure}
\centering
\begin{subfigure}
  \centering
  \includegraphics[width=0.4\linewidth]{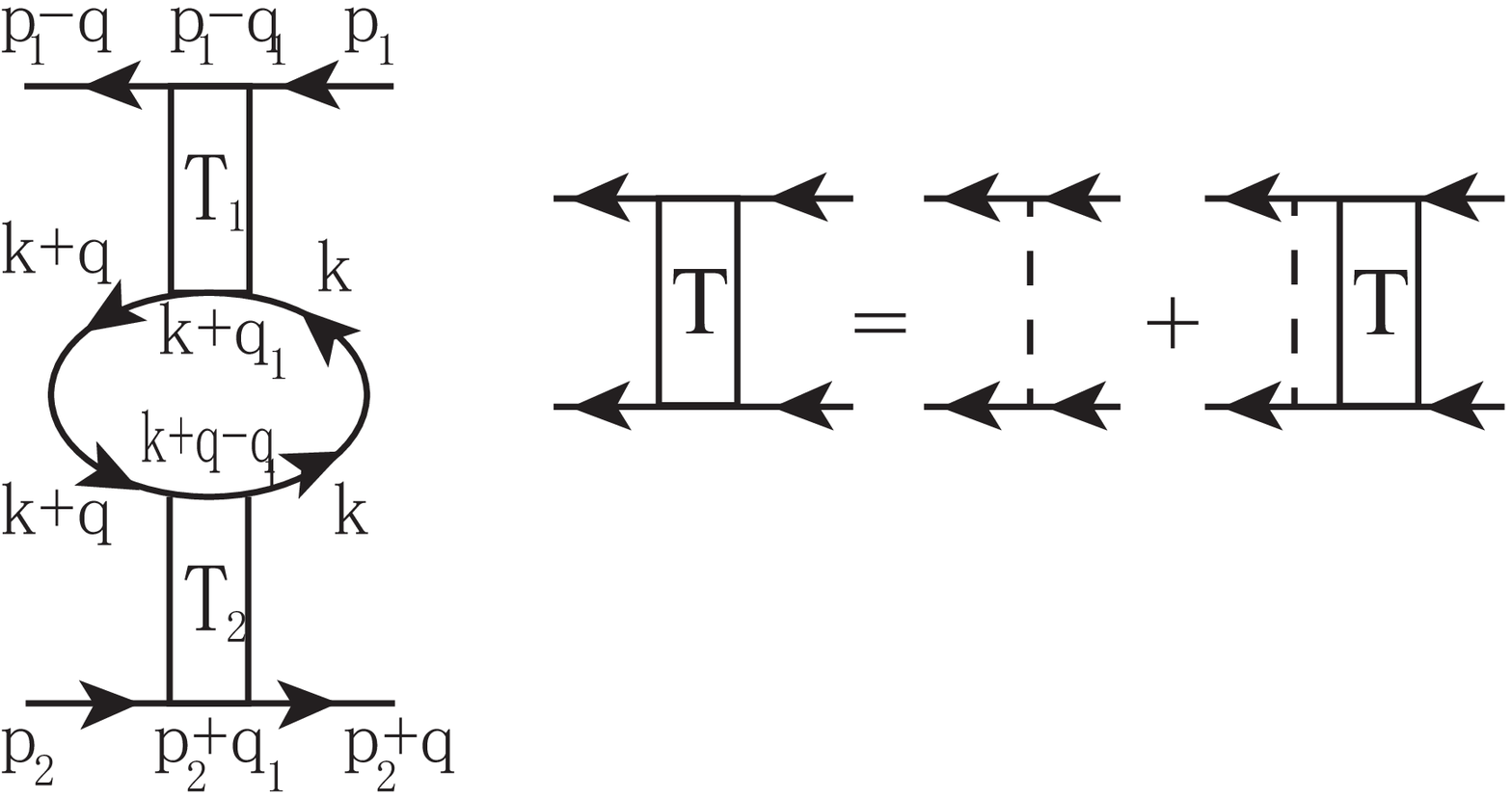}
\end{subfigure}
\caption{The diagram of bipolaron.
}
\end{figure}

\begin{figure}
\centering
\begin{subfigure}
  \centering
  \includegraphics[width=0.4\linewidth]{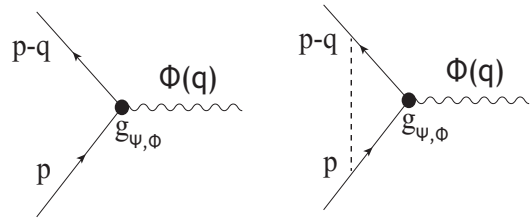}
\end{subfigure}
\caption{
The Yukawa coupling,
which consist of the two fermion propagators and one boson propagator 
and a coupling between them in the vertex.
The dashed line in the right-hand-side panel is the four fermions interaction,
e.g., the disorder couplings due to Coulomb interaction, which will not be considered in this paper.
}
\end{figure}

\clearpage
\begin{figure}
\centering
\begin{subfigure}
  \centering
  \includegraphics[width=0.4\linewidth]{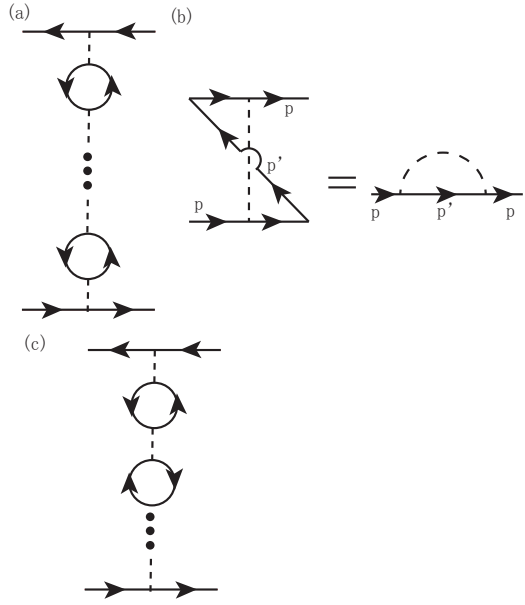}
\end{subfigure}
\caption{
(a) The diagram of off-diagonal self-energy Eq.(32) containing an infinite number of bosonic bubbles in Eliashberg-Migdal approximation
(to first order in coupling).
We also shown in (b) the contribution of mean field contribution to anomalous self-energy,
which, to first order of coupling,
it's equivalent to a single-particle propagator with a side-interaction.
(c) shows another type of boson propagator as described by Eq.(33).
}
\end{figure}

\begin{figure}
\centering
\begin{subfigure}
  \centering
  \includegraphics[width=0.4\linewidth]{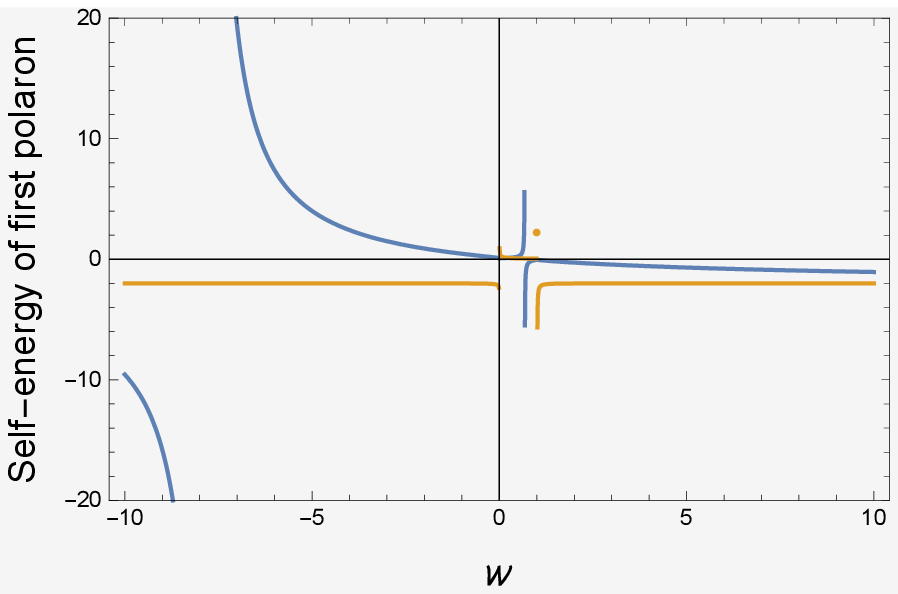}
\end{subfigure}
\begin{subfigure}
  \centering
  \includegraphics[width=0.4\linewidth]{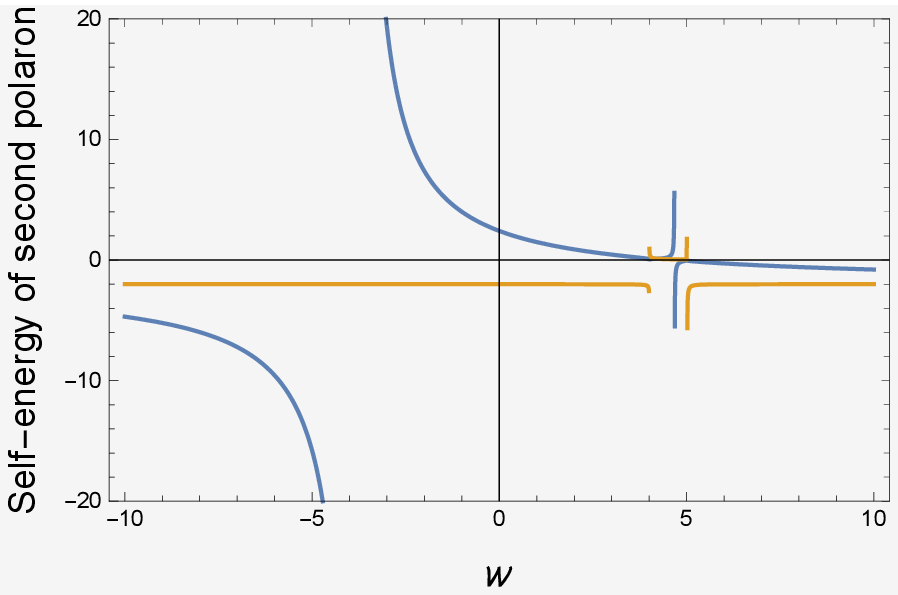}
\end{subfigure}
\begin{subfigure}
  \centering
  \includegraphics[width=0.4\linewidth]{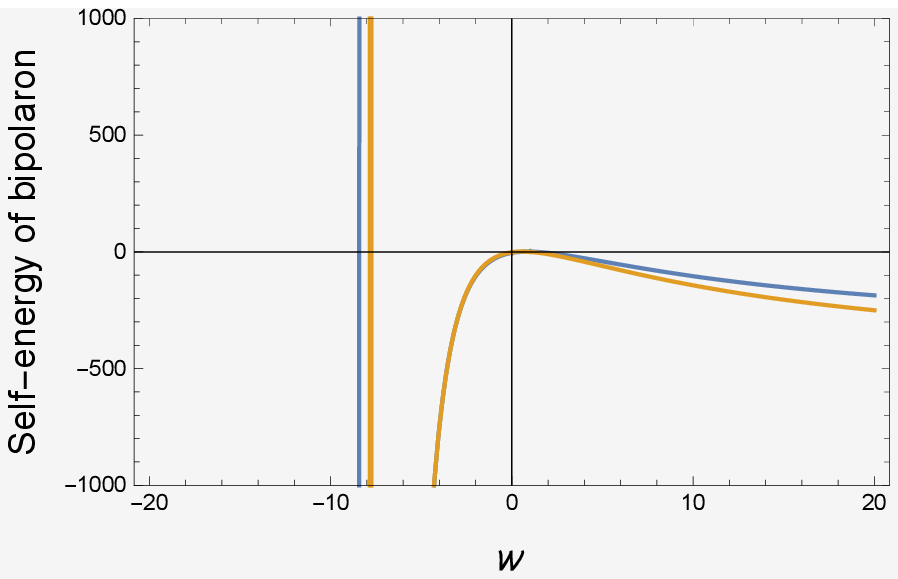}
\end{subfigure}
\begin{subfigure}
  \centering
  \includegraphics[width=0.4\linewidth]{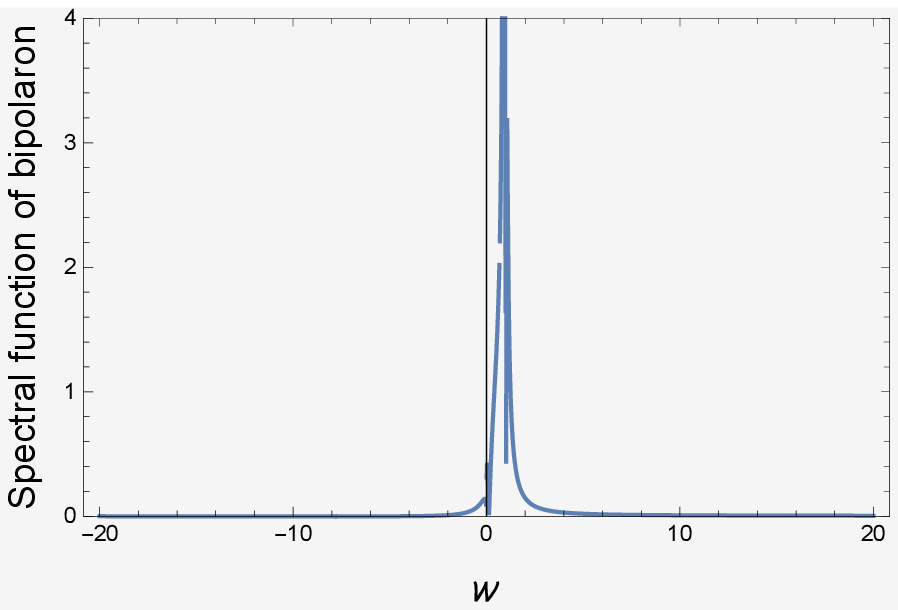}
\end{subfigure}
\caption{
The self-energies of single-polarons and the bipolaron at the framework of zero center-of-mass momentum ($Q=0$).
In this diagram we set $\Omega=0$ for simplicity.
In this simulation,
the UV curoff are setted as $\Lambda_{q}=\Lambda_{k}=2\Lambda_{q_{1}}$.
The blue and yellow lines correspond to the real part and imaginary part, respectively.
}
\end{figure}

\clearpage
\begin{figure}
\centering
\begin{subfigure}
  \centering
  \includegraphics[width=0.4\linewidth]{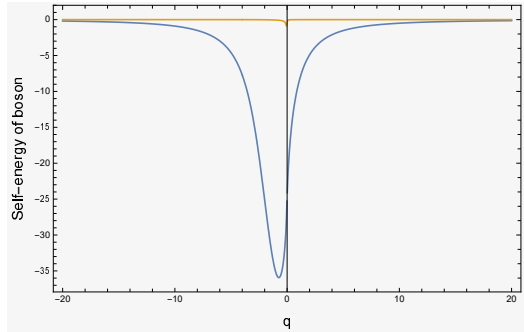}
\end{subfigure}
\caption{
Bosonic self-energy at zero temperature as presented in Eq.().}
\end{figure}

\begin{figure}
\centering
\begin{subfigure}
  \centering
  \includegraphics[width=0.4\linewidth]{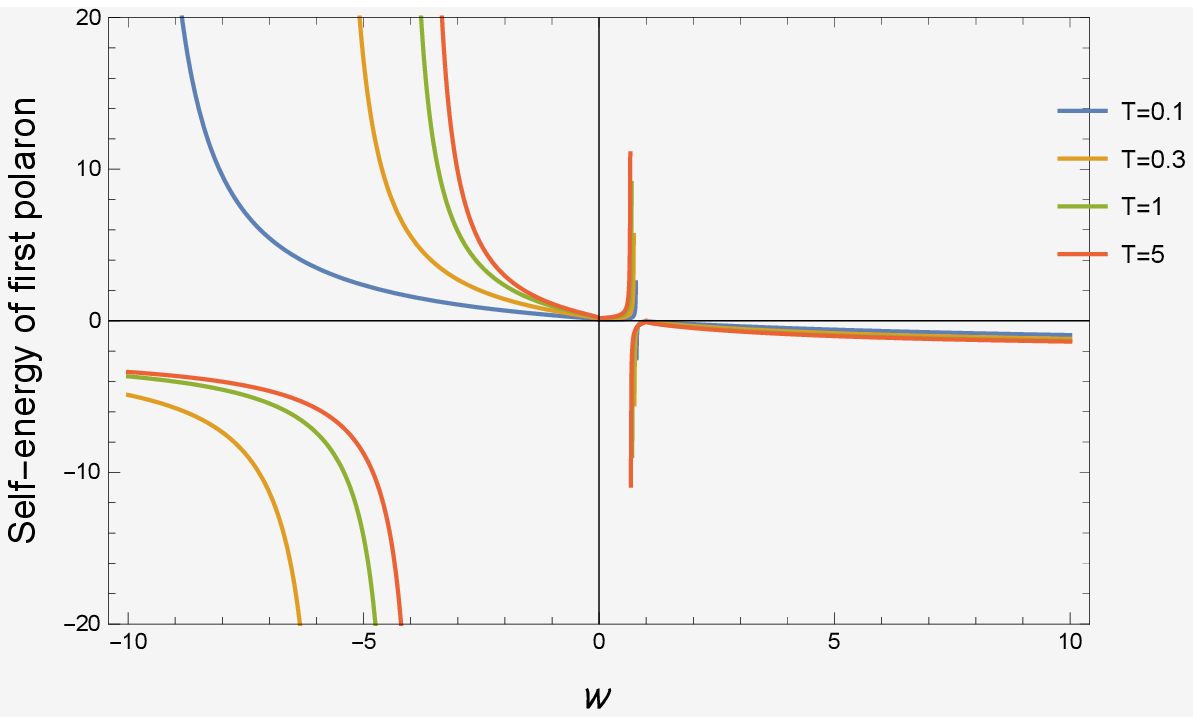}
\end{subfigure}
\begin{subfigure}
  \centering
  \includegraphics[width=0.4\linewidth]{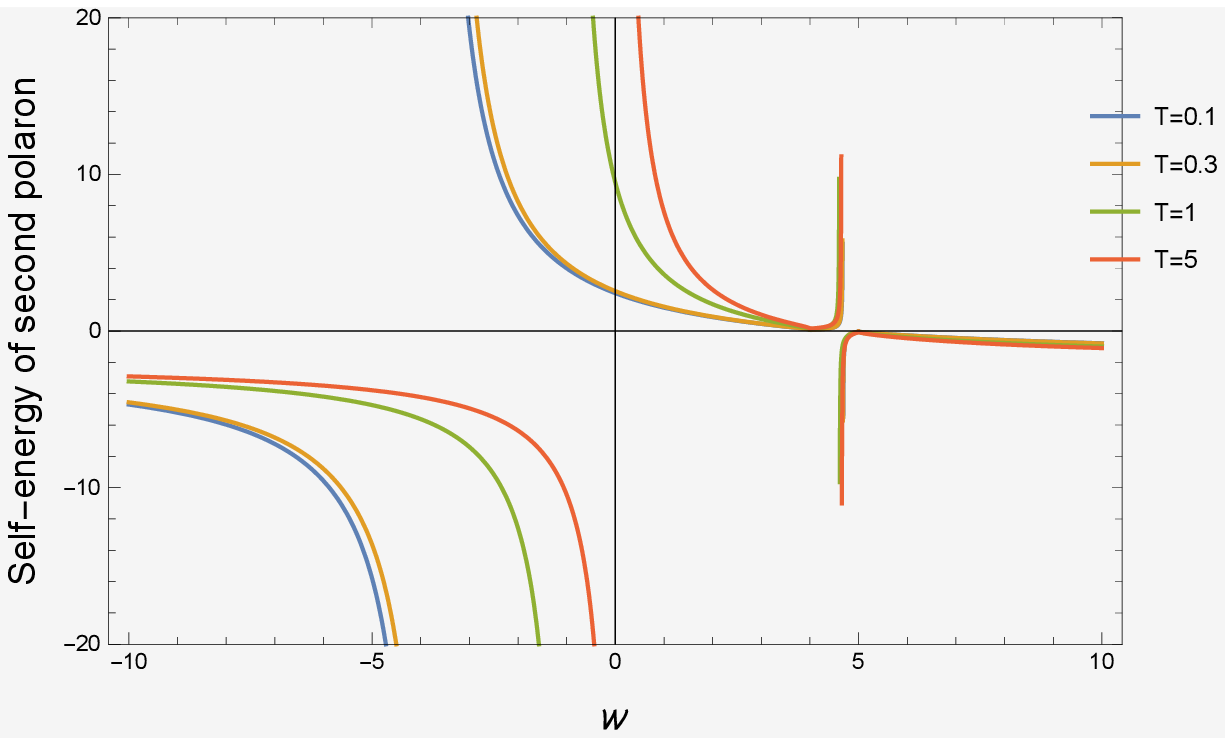}
\end{subfigure}
\begin{subfigure}
  \centering
  \includegraphics[width=0.4\linewidth]{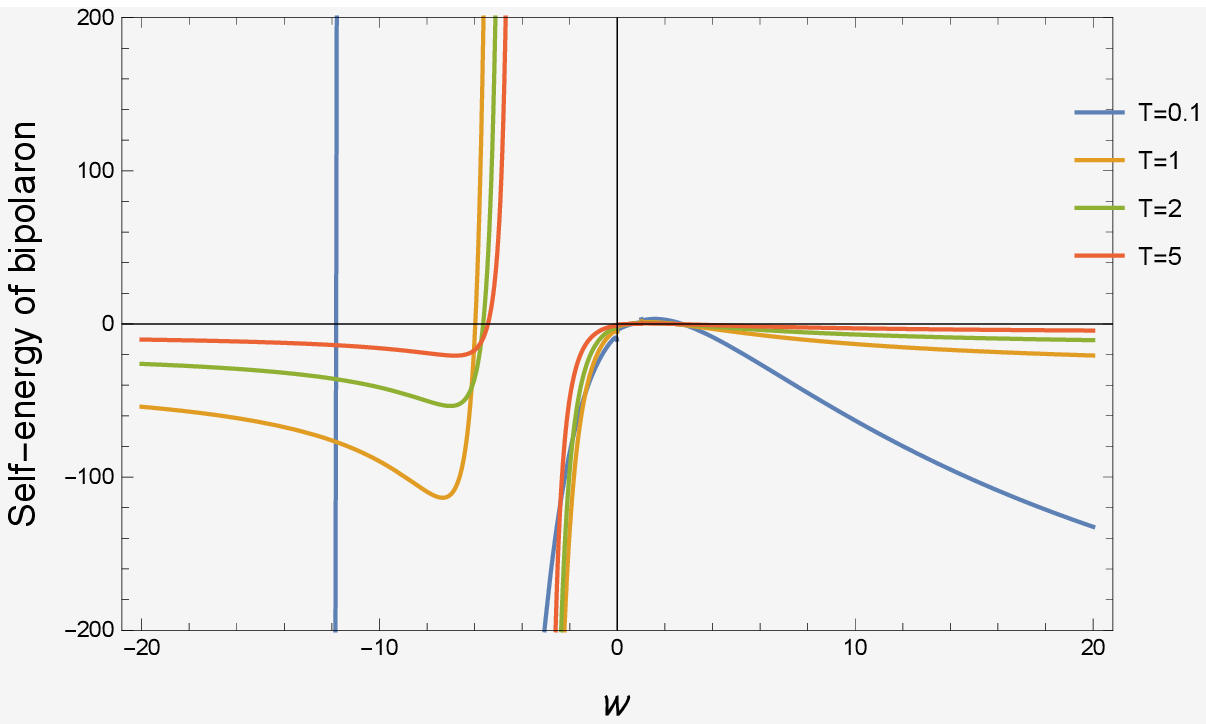}
\end{subfigure}
\begin{subfigure}
  \centering
  \includegraphics[width=0.4\linewidth]{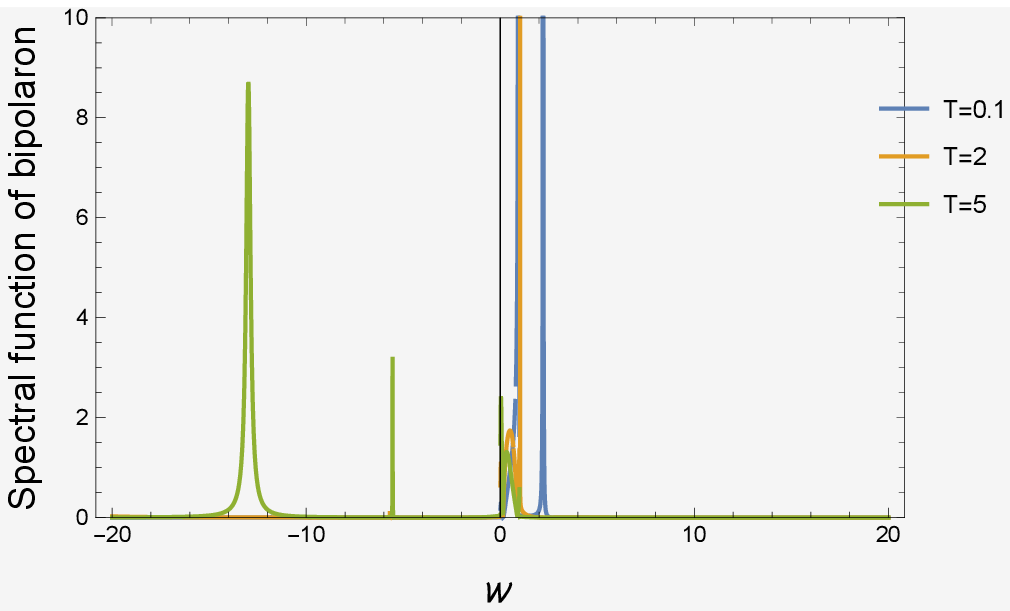}
\end{subfigure}
\caption{
Similar to Fig.(4) but for finite temperature.
}
\end{figure}

\clearpage
\begin{figure}
\centering
\begin{subfigure}
  \centering
  \includegraphics[width=0.4\linewidth]{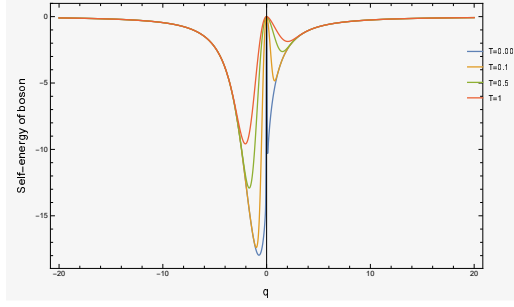}
\end{subfigure}
\caption{
Bosonic self-energy at zero temperature.}
\end{figure}

\begin{figure}
\centering
\begin{subfigure}
  \centering
  \includegraphics[width=0.4\linewidth]{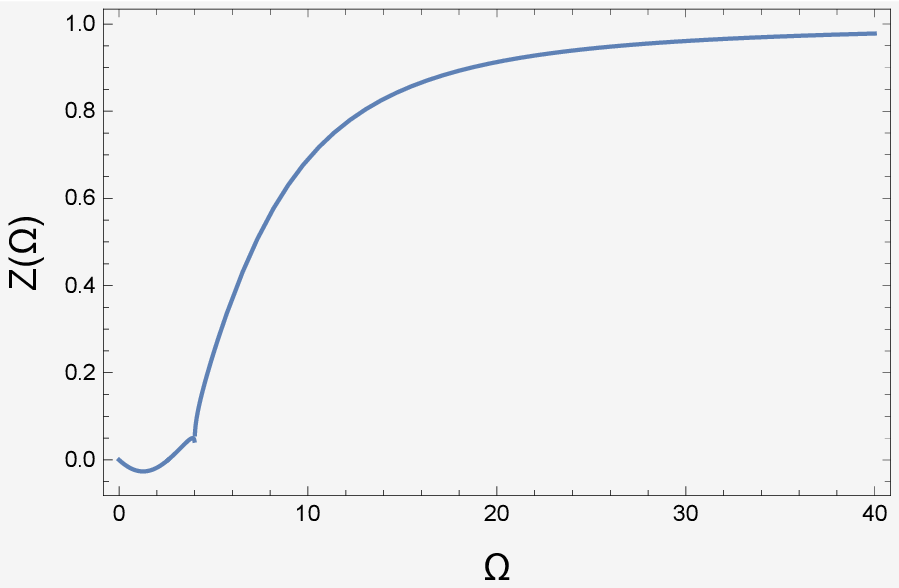}
\end{subfigure}
\begin{subfigure}
  \centering
  \includegraphics[width=0.4\linewidth]{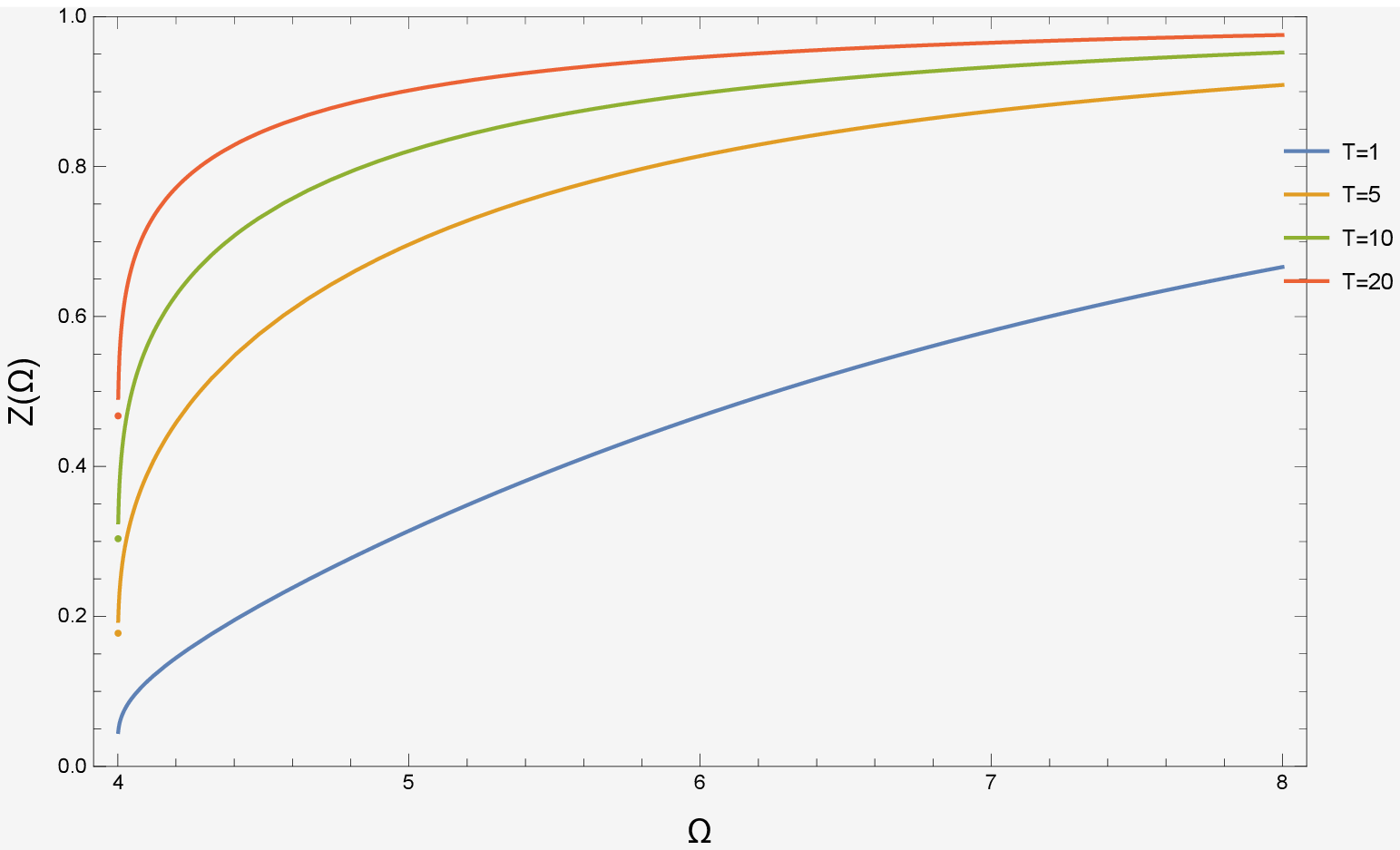}
\end{subfigure}
\caption{
(left) The quasiparticle residue at zero temperature as described by Eq.(9).
(right) The quasiparticle residue at finite temperature.
}
\end{figure}

\begin{figure}
\centering
\begin{subfigure}
  \centering
  \includegraphics[width=0.4\linewidth]{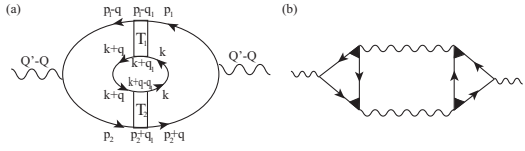}
\end{subfigure}
\caption{(a) Aslamazov-Larkin-type
diagram of correlation function.
(b) The standard Aslamazov-Larkin diagram with four
 shaded vertices representing the coherent scattering of two particles.
The wavy lines represent the vertices (with external momentum) attached to the fermion propagators.
}
\end{figure}

\begin{figure}
\centering
\begin{subfigure}
  \centering
  \includegraphics[width=0.4\linewidth]{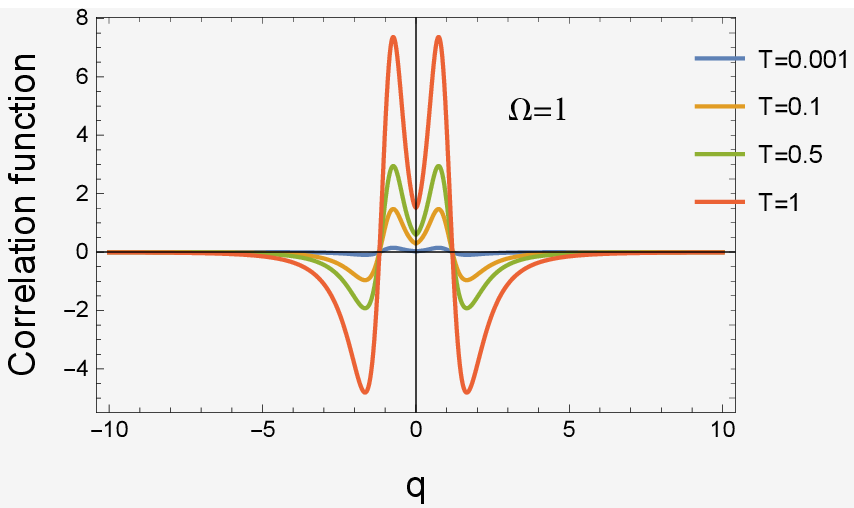}
\end{subfigure}
\begin{subfigure}
  \centering
  \includegraphics[width=0.4\linewidth]{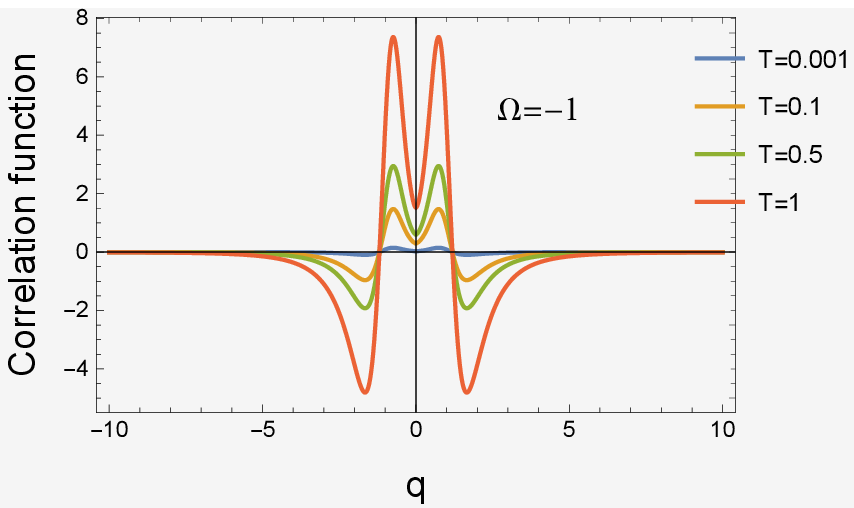}
\end{subfigure}
\begin{subfigure}
  \centering
  \includegraphics[width=0.4\linewidth]{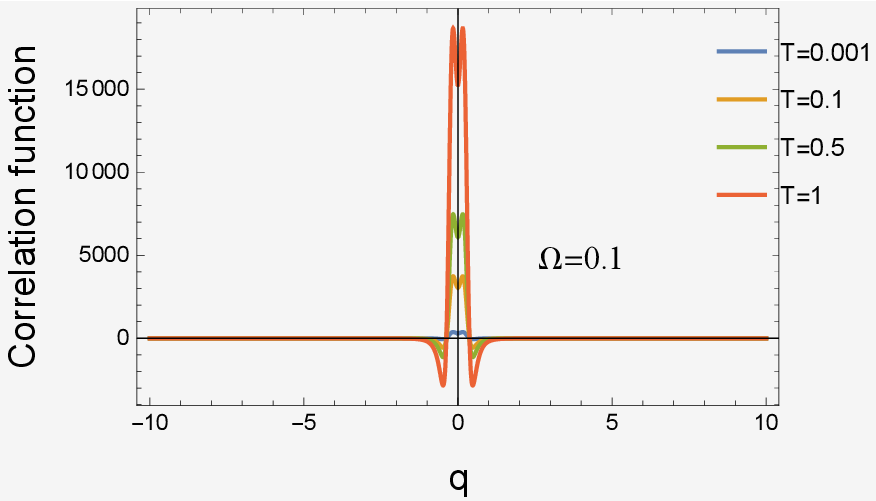}
\end{subfigure}
\caption{Real part of the correlation function $\chi(q,\Omega)$.
}
\end{figure}

\clearpage
\begin{figure}
\centering
\begin{subfigure}
  \centering
  \includegraphics[width=0.4\linewidth]{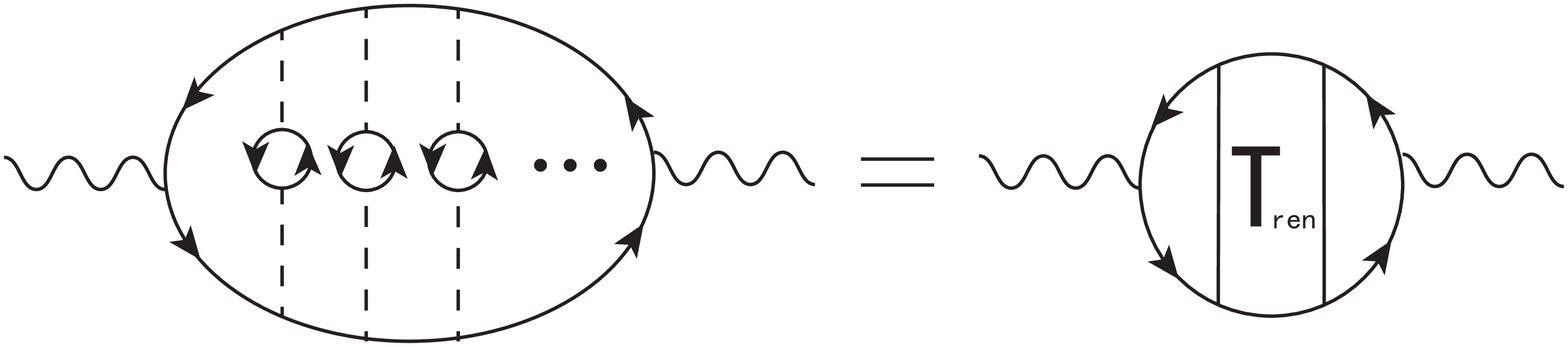}
\end{subfigure}
\caption{The AL-type diagram with the renormalized $T$-matrix defined in Eq.(73).
}
\end{figure}

\begin{figure}
\centering
\begin{subfigure}
  \centering
  \includegraphics[width=0.4\linewidth]{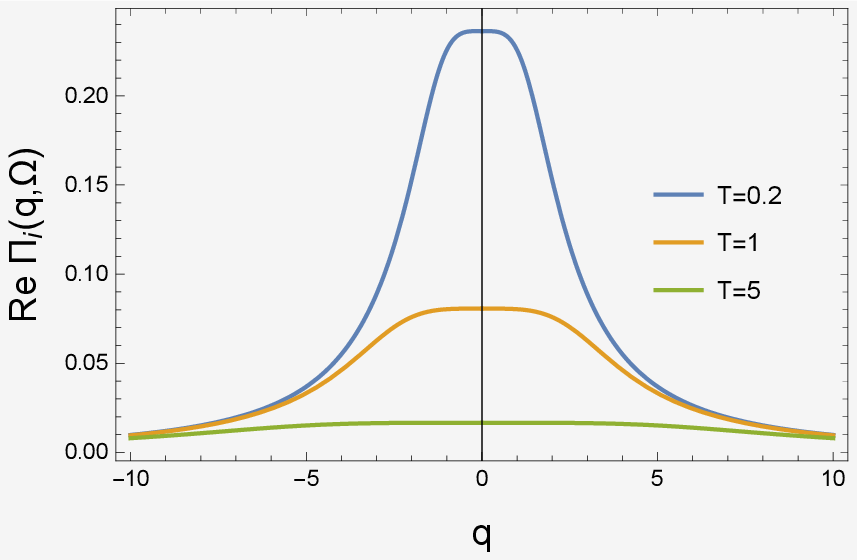}
\end{subfigure}
\begin{subfigure}
  \centering
  \includegraphics[width=0.4\linewidth]{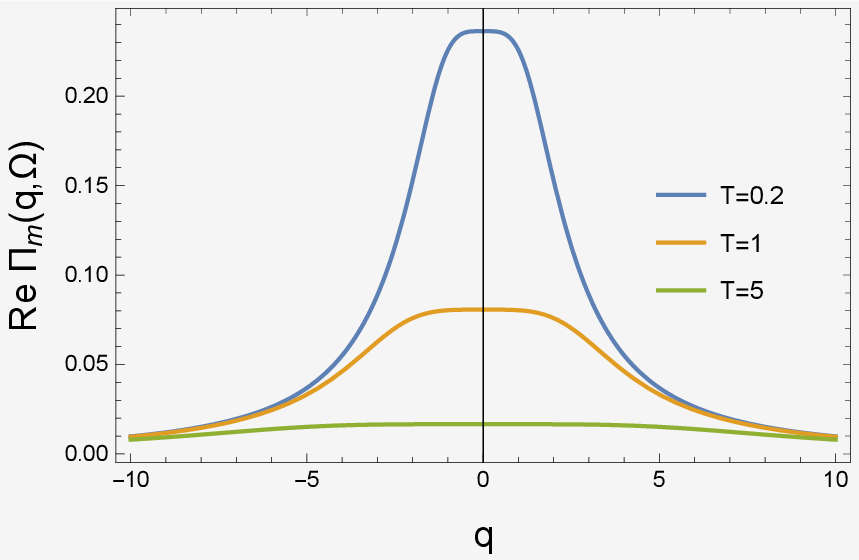}
\end{subfigure}
\begin{subfigure}
  \centering
  \includegraphics[width=0.4\linewidth]{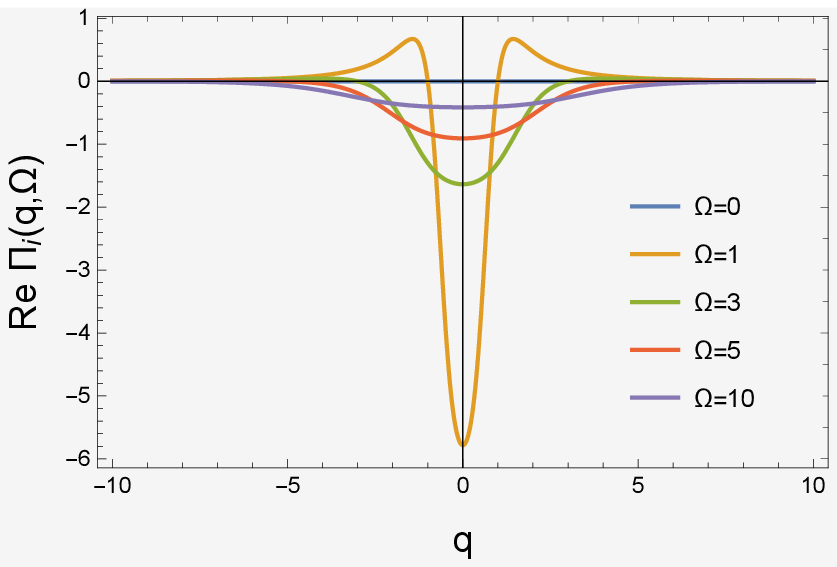}
\end{subfigure}
\begin{subfigure}
  \centering
  \includegraphics[width=0.4\linewidth]{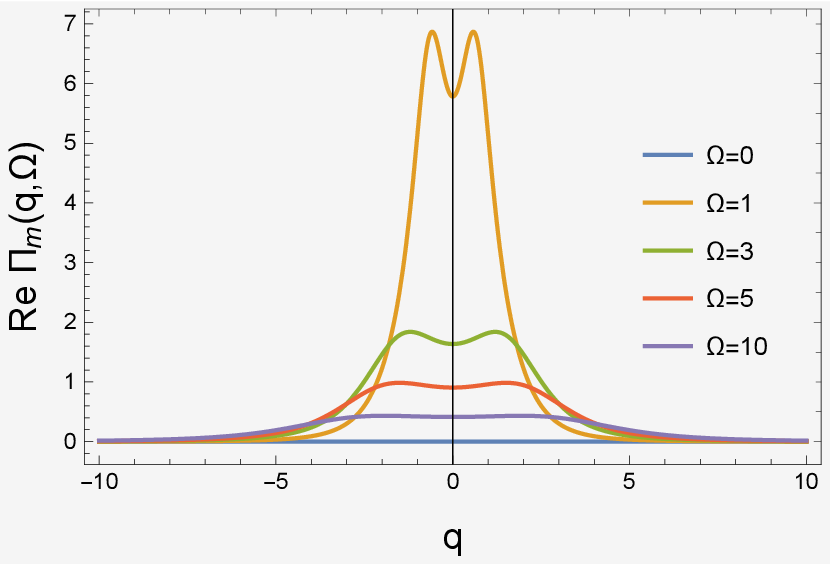}
\end{subfigure}
\caption{The real part of two-particle correlations defined in Eq.(72).
The upper panels are at finite-temperature and with $\Omega=0$,
where we approximate temperature-dependent correlations as independent of the very small quantity $\delta q$
since $\Pi(T,q)=\Pi_{0}(T)+O(\delta q^{2})$.
And under this approximation, ${\rm Re}\Pi_{i}={\rm Re}\Pi_{m}$.
The lower panels are at zero-temperature.
}
\end{figure}

\begin{figure}
\centering
\begin{subfigure}
  \centering
  \includegraphics[width=0.4\linewidth]{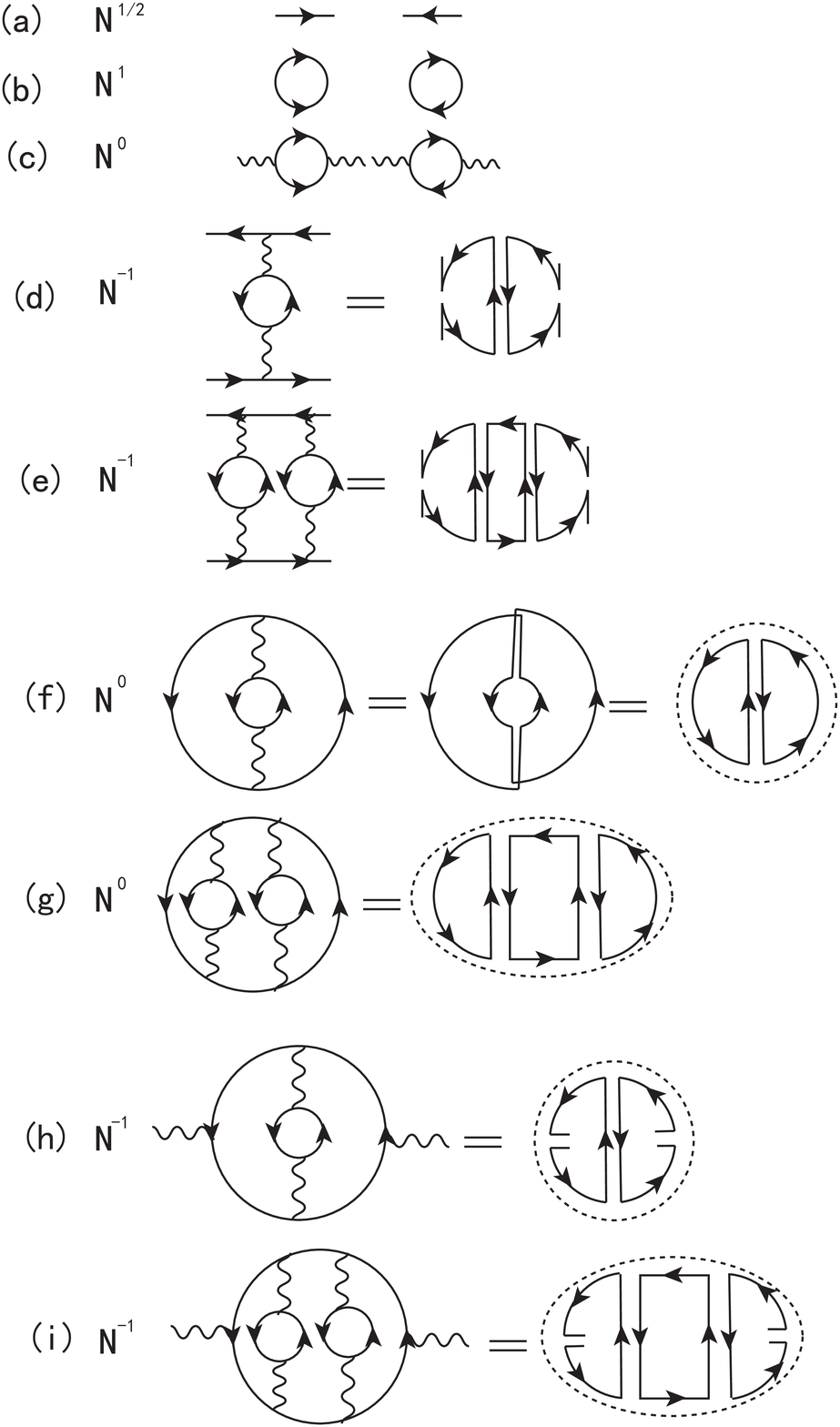}
\end{subfigure}
\caption{
The order of Feynman diagrams to leading order in $1/N$. 
}
\end{figure}

\begin{figure}
\centering
\begin{subfigure}
  \centering
  \includegraphics[width=0.4\linewidth]{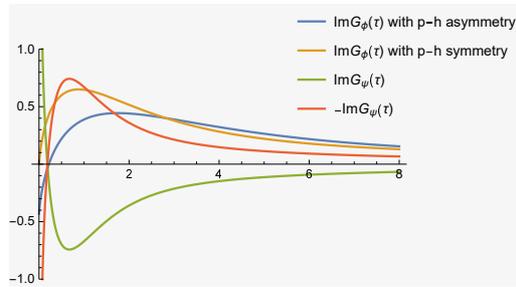}
\end{subfigure}
\caption{
The imaginary part of fermion field and boson field propagators in imaginary time domain.
}
\end{figure}

\begin{figure}
\centering
\begin{subfigure}
  \centering
  \includegraphics[width=0.4\linewidth]{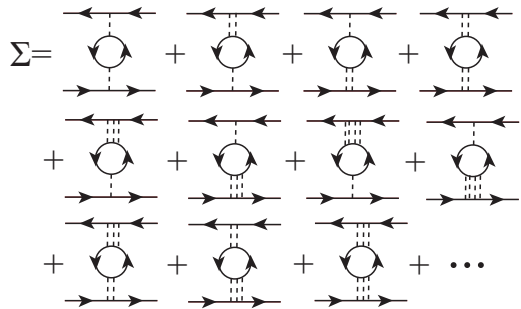}
\end{subfigure}
\caption{Diagram of Eq.(89).
}
\end{figure}

\end{document}